\pdfoutput=1
\documentclass[12pt]{article}

\usepackage{amsmath,amssymb,bm}
\usepackage{graphicx}
\usepackage{booktabs}
\usepackage{microtype}
\usepackage{float}
\usepackage{placeins}
\usepackage{pgfplots}
\pgfplotsset{compat=1.18}
\usepgfplotslibrary{groupplots}

\usepackage{braket}
\usepackage{geometry}
\usepackage[
colorlinks=true,
linkcolor=blue,
citecolor=blue,
urlcolor=blue,
breaklinks=true
]{hyperref}

\title{\textbf{The Fock representation of the Coulomb interaction.\\ I. Two-body problem}}
\author{M.~M.~Nishonov\thanks{Email: m.nishonov@nuu.uz}\\
	National University of Uzbekistan, 100174 Tashkent, Uzbekistan}
\date{}

\begin{document}
\maketitle

\begin{abstract} In the momentum-space few-body theory of light nuclei, where the Faddeev equations with separable two-body input close on one-variable amplitudes, the unscreened Coulomb interaction is exactly separable by the stereographic projection of Fock, with analytic form factors and strengths, and the truncation of the expansion converges at negative energies. It is derived here as an operator identity for the potential and added, without screening or fitting, as a block of separable terms to the separable nuclear input of a charged pair. The representation is compared with the screening--renormalization method and with Coulomb-distorted form factors on two-body benchmarks: the off-shell $t$~matrix, the Coulomb displacement of the $\alpha p$ Pauli-forbidden state, and the Coulomb-modified phases of the $pp$, $\alpha p$ and $\alpha\alpha$ pairs. The Feshbach--Schur projection of the Pauli-forbidden state is shown to require the Coulomb-dressed eigenstate, which the Fock block provides within the same separable problem. At positive energy a finite truncation is a screened Coulomb potential with an analytic, momentum-dependent radius.
\end{abstract}

\section{Introduction}
\label{sec:intro}

The Coulomb interaction enters the momentum-space few-body theory of nuclei, the Faddeev equations for the three-nucleon system and for the light nuclei which the cluster model describes as three bodies, $^{6}$He, $^{6}$Li, $^{9}$Be, $^{9}$B and $^{12}$C, through its matrix element
\begin{equation*}
\braket{\bm p|V_{C}|\bm p'}=\frac{Z_{1}Z_{2}\,e^{2}}{2\pi^{2}\,|\bm p-\bm p'|^{2}},
\end{equation*}
which is singular at zero momentum transfer; the singularity makes the kernel of the two-body Lippmann--Schwinger equation non-compact and passes with the two-body $t$~matrix into the kernel of the Faddeev equations. The established treatment screens it, solves the equations at a finite screening radius and removes the screening by the renormalization of Ref.~\cite{Alt1978}, as in the proton--deuteron calculations of Refs.~\cite{Alt2002,Deltuva2005}. The limit is numerical and every result has a systematic error from the screening radius. Which amplitudes require the renormalization is still debated: Deltuva~\cite{Deltuva2024} finds it necessary for the elastic scattering amplitude, where the screened treatment of Wita{\l}a \emph{et al.}~\cite{Witala2009} had omitted it, and those authors maintain their view in response~\cite{Witala2024}. For Faddeev calculations with separable two-body input the screened tail has a second drawback: it is not separable, and adding it destroys the separable structure on which the efficiency of such calculations rests. The alternative that is natural for separable potentials, dressing the nuclear form factors with the screened Coulomb wave operator, includes the Coulomb--nuclear interference in the form factors and is examined here beside the screening method.

This paper introduces the Coulomb force into a separable kernel in a third way, by interpreting Fock's $O(4)$ symmetry~\cite{Fock1935} as an operator identity for the potential. After the stereographic projection onto the three-sphere the Coulomb kernel is a convolution, diagonal in the $S^{3}$ harmonics; the unscreened Coulomb potential is therefore exactly separable at negative energies, with analytic form factors and analytic strengths. A bound-state Faddeev kernel samples every pair interaction only at negative pair energies (Ref.~\cite{FockII}), so the Coulomb force is added to the separable nuclear input as a further block of separable terms, without a screening radius, a fit or a limiting procedure, and the truncation converges. That the same truncation cannot represent the on-shell singularity may be why the representation has not been used for few-body kernels; here that limitation is quantified and then exploited. The second element is the strict elimination of Pauli-forbidden states of Ref.~\cite{Nishonov2026a}, the Schur complement over a forbidden block of the same separable basis, whose only condition is that the projected vector be an eigenvector of the pair Hamiltonian that actually enters the calculation. With the Coulomb force present that vector is the Coulomb-\emph{dressed} forbidden state, and the Fock block makes it available inside the same separable problem; the condition is tested by comparing the two-cluster phase shifts obtained with the eigenstate and with the oscillator function of the orthogonality-condition model~\cite{Saito1969}. The third element extends the representation to positive energy: the two-body Coulomb-modified phase is obtained with the singularity treated exactly, and a finite truncation is shown to be a screened Coulomb potential with an analytic, momentum-dependent radius, whose renormalized phase converges to the exact one. Reference~\cite{FockII} uses the same two-body input in the three-body kernel and gives the bound states of the three-nucleon and three-cluster nuclei.

Fock's method has been applied to the two-body Coulomb $t$~matrix itself in Refs.~\cite{Kharchenko2018,Kharchenko2017a,Kharchenko2017b,Kharchenko2019}, where analytic partial-wave Coulomb $t$~matrices were obtained at negative energies for integer values of the Coulomb parameter, the bound-state energies of the attractive problem and their repulsive counterparts; the present representation holds at every negative energy, is separable with analytic strengths, and joins the separable nuclear input. The exact Coulomb $t$~matrix entered a momentum-space three-body bound-state calculation once before, in the trinucleon model of Kok and van Haeringen for the $^{3}$He Coulomb energy~\cite{vanHaeringen1982}; here the potential itself is separable and the kernel is unchanged. Separable series for the off-shell Coulomb $t$~matrix at fixed energy were introduced in atomic physics by Chen and Ishihara~\cite{ChenIshihara1969}. The Coulomb--Sturmian expansion of Papp and Plessas~\cite{PappPlessas1996,Papp1997} uses, up to normalization and energy parametrization, the Fock modes of this paper, but expands the potentials and Green's operators in that basis, whereas here the same functions form an identity for the potential with analytic strengths and the nuclear part is not expanded. The Coulomb-basis Faddeev formulation of Mukhamedzhanov, Eremenko and Sattarov~\cite{Mukhamedzhanov2012}, realized for optical potentials by Hlophe, Elster, Nunes and collaborators~\cite{Hlophe2014,Eremenko2015,Hlophe2017} and reviewed in Ref.~\cite{ElsterNTSE2014}, is the standard form of the distorted-form-factor method for $(d,p)$ reactions on heavy nuclei, with an unscreened regularized folding; it differs from the screened M{\o}ller dressing compared below. What is new here, relative to all of these, is the Coulomb block with analytic strengths inside an otherwise unchanged separable kernel, and the projection, within that same problem, of the Coulomb-dressed forbidden state which the eigenstate condition requires. The practical advantage is a bound-state three-body calculation without screening radius or limiting procedure, the price being some twenty additional separable terms per charged pair.

The paper is organized as follows. In Sec.~\ref{sec:fock} the representation is introduced, the extended separable problem of a charged pair is formulated, and the limit of validity at positive energy is established. The eigenstate condition and the Coulomb-dressed forbidden states are derived in Sec.~\ref{sec:exact}. Section~\ref{sec:routes} formulates the screening method and the Coulomb-distorted form factors within the separable formalism and compares them with the Fock representation on four two-body benchmarks; the eigenstate condition is then tested in Sec.~\ref{sec:prescription} on the two-cluster phase shifts, with the oscillator function substituted for the dressed eigenstate. The extension to positive energies, the Coulomb-modified phase shift and the region of positive pair energies above the three-body breakup threshold, is the subject of Sec.~\ref{sec:scattering}, and the relation to the separable kernels of effective field theory is discussed briefly in the summary (Sec.~\ref{sec:summary}). The derivation of the representation, the oscillator forbidden states and the parameters of the separable potentials are given in the appendices.

\section{The Fock representation and the limit of its validity}
\label{sec:fock}

The nuclear input of all calculations in this paper is separable in the sense of Ernst, Shakin and Thaler~\cite{EST1973}. Each pair interaction is a finite sum of separable terms, and its $t$~matrix and propagator matrix have the same finite-rank form,
\begin{equation}
V_{s} \;=\; \sum_{ij}\ket{\chi_{i}}\,\Lambda_{ij}\,\bra{\chi_{j}},
\qquad
t_{s}(z) \;=\; \sum_{ij}\ket{\chi_{i}}\,\tau_{ij}(z)\,\bra{\chi_{j}},
\label{eq:est}
\end{equation}
the $t$~matrix having the same form factors as the potential and all the energy dependence in the matrix
\begin{equation}
\bm\tau(z)=\bigl[\bm\Lambda^{-1}-\bm{\mathcal D}(z)\bigr]^{-1},
\qquad
\mathcal D_{ij}(z)=\braket{\chi_{i}|G_{0}(z)|\chi_{j}},
\label{eq:tau}
\end{equation}
with $i,j=1,\dots,N_{r}$, $N_{r}$ the rank, $\bm\Lambda$ the coupling matrix, $G_{0}(z)=(z-H_{0})^{-1}$ the free resolvent at the pair energy $z$, the propagator matrix being the radial integral~(\ref{eq:measure}) of Appendix~\ref{app:derivation}, $\bm\tau(z)$ the pair $t$~matrix in the space of the form factors, and with the analytic form-factor ansatz of the separable potentials constructed here,
\begin{equation}
\chi^{(L)}_{i}(p) \;=\; \frac{p^{L}}{(p^{2}+\beta^{2})^{M}}
\sum_{n=0}^{n_{c}} a^{(L)}_{in}\,P_{n}(x),
\qquad
x=\frac{\beta^{2}-p^{2}}{\beta^{2}+p^{2}},
\label{eq:estff}
\end{equation}
where $i$ labels the support points, $L$ is the partial wave of the channel (the superscript is dropped where the partial wave is fixed), $M=1+\lfloor L/2\rfloor$ the integer power chosen in the construction of the separable potentials, $P_{n}$ are the Legendre polynomials, $\beta$ the pole scale of the channel, $a^{(L)}_{in}$ fitted coefficients and $n_{c}$ the order of the expansion, $22$ for the $NN$, $10$ for the $\alpha N$ and $12$ for the $\alpha\alpha$ potentials (Table~1 of Ref.~\cite{NishonovValidation}). In a coupled channel each $\chi_{i}$ has a component of the form~(\ref{eq:estff}) in each of the two waves, with the $L$ and $M$ of that wave. Because the representation is intended as the input of the momentum-space three-body equations of Ref.~\cite{FockII}, the propagator integrals, the pair $\tau$~matrices and the exchange elements of their kernel must all be built from this family of functions.

The unscreened Coulomb interaction belongs to the same family. Its $L=0$ partial wave has the separable form
\begin{equation}
V_{C,0}(p,p') \;=\; \sum_{n=0}^{\infty}
\Phi_{n}(p)\,\lambda^{C}_{n}\,\Phi_{n}(p'),
\qquad
\Phi_{n}(p) = \frac{C^{(1)}_{n}(x_{C})}{\beta_{C}^{2}+p^{2}},
\qquad
x_{C} = \frac{\beta_{C}^{2}-p^{2}}{\beta_{C}^{2}+p^{2}},
\label{eq:fock}
\end{equation}
with
\begin{equation}
\lambda^{C}_{n} \;=\; \frac{8\, Z_{1}Z_{2}\,e^{2}\, \beta_{C}^{2}}
{\pi\,(n+1)},
\label{eq:lamn}
\end{equation}
where $C^{(1)}_{n}$ are the Gegenbauer (Chebyshev--$U$) polynomials and $\beta_{C}$ a scale parameter: the same stereographic variable, $x_{C}$ being the variable $x$ of~(\ref{eq:estff}) at $\beta=\beta_{C}$, the same pole factors, orthogonal polynomials of the same class, and no fitted quantity, since the strengths~(\ref{eq:lamn}) are analytic. The result follows from the stereographic projection of Fock, under which the Coulomb kernel is a convolution on the three-sphere and diagonal in its harmonics; the derivation, from the expression of the momentum transfer through the chord of the four-dimensional sphere, Eq.~(\ref{eq:chord}), to the normalization of the strengths, is in Appendix~\ref{app:derivation}. A general partial wave has the same form with
\begin{equation}
\Phi_{nL}(p) \;=\; \frac{p^{L}\,C^{(L+1)}_{n}(x_{C})}{(\beta_{C}^{2}+p^{2})^{L+1}},
\label{eq:fockL}
\end{equation}
and strengths
\begin{equation}
\lambda_{nL} \;=\; \frac{2^{4L+3}\,Z_{1}Z_{2}\,e^{2}\,\beta_{C}^{2L+2}\,n!\,(L!)^{2}}
{\pi\,(n+2L+1)!},
\label{eq:lamnL}
\end{equation}
which reduce to Eq.~(\ref{eq:lamn}) for $L=0$ and fall as $1/[(n+1)(n+2)\cdots(n+2L+1)]$, the $1/(n+L+1)$ of the $S^{3}$ eigenvalue combined with the normalization of the harmonics (Appendix~\ref{app:derivation}). The scale $\beta_{C}$ determines the rate of convergence of the expansion, not its limit; $\beta_{C}=0.8$~fm$^{-1}$ is used for $\alpha p$ and $1.0$~fm$^{-1}$ for $pp$ and $\alpha\alpha$ in the two-body calculations of this paper.

\subsection{The separable nuclear potentials}
\label{sec:tables}

The nuclear input is the set of separable potentials of Ref.~\cite{NishonovValidation}, constructed in the Ernst--Shakin--Thaler form~\cite{EST1973}. For support states $\ket{\psi_{i}}$ of the original interaction $V$ (the interaction being represented) at energies $E_{i}$, $i=1,\dots,N_{r}$, with $N_{r}$ the rank,
\begin{equation}
\chi_{i}=V\ket{\psi_{i}},\qquad
\bigl[\bm\Lambda^{-1}\bigr]_{ij}=\braket{\psi_{i}|V|\psi_{j}},
\label{eq:estdef}
\end{equation}
so that the separable $t$~matrix of Eq.~(\ref{eq:est}) is exact half on shell (one momentum argument on the energy shell) at every support. Every bound state of the original potential is a support point. The other support points are half-shell columns of the standing-wave $K$~matrix at chosen energies of both signs, $\chi_{i}(p)=K(p,k_{i};E_{i})$, which coincide with the half-shell $t$~matrix on shell: positive-energy columns suppress the spurious deep states that a set of support points lying entirely at negative energy can produce, and negative-energy columns, $E_{i}<0$ at chosen momenta $k_{i}$, cover the off-shell region the kernel samples and remain smooth where scattering states with a long-range force do not. The exact half-shell form factors are then fitted to the analytic family~(\ref{eq:estff}) with the integer $M=1+\lfloor L/2\rfloor$ ($M=1$ for $S$ and $P$ waves, $2$ for $D$ and $F$, $3$ for $G$ and $H$), raised by one for the $^{3}F_{3}$ and $^{3}F_{4}$--$^{3}H_{4}$ potentials (Ref.~\cite{NishonovValidation}); this choice minimizes the Legendre coefficients. The coupling matrix is then recomputed from the fitted form factors and symmetrized, so that the form factors and the coupling matrix that enter the three-body equations are consistent to the accuracy of the fit. The Pauli-forbidden states are the bound states of the fitted separable potential itself, represented in the two-pole form
\begin{equation}
\phi_{fL}(p)\propto\frac{p^{L}\sum_{n=0}^{n_{c}}A^{(f)}_{n}P_{n}(x)}{(p^{2}+\beta^{2})(p^{2}+\gamma_{f}^{2})},
\qquad \gamma_{f}=\sqrt{2\mu|E_{f}|},
\label{eq:fbff}
\end{equation}
with $f$ labelling the state and $L$ its partial wave, $\gamma_{f}$ the bound-state momentum and $x$ the variable of Eq.~(\ref{eq:estff}). For the channels with $M=1$ this is the exact bound state of the separable potential, the $A^{(f)}_{n}$ being the coefficients of the bound-state vector in the form factors~(\ref{eq:estff}), reproduced exactly by the two-pole form, the pole of the dressed $\alpha p$ state in close agreement with the converged Fock eigenvalue (Table~\ref{tab:pairs2}, Sec.~\ref{sec:prescription}); for $M=2$ the single pole at $\beta$ is a refit of the exact form, which it reproduces accurately, for the $\alpha\alpha$ $0d$ state. The quality of a separable potential is judged by its off-shell accuracy at the negative pair energies which the three-body bound-state equations sample,
\begin{equation}
\Xi(E)=
\frac{\lVert t_{\rm sep}(p,p';E)-t_{\rm orig}(p,p';E)\rVert_{w}}
     {\lVert t_{\rm orig}(p,p';E)\rVert_{w}},
\label{eq:eps}
\end{equation}
with $p,p'\le3$~fm$^{-1}$, $E$ from $-5$ to $-200$~MeV and the $p^{2}$ weight of the kernel; the support points of the $NN$ and $\alpha\alpha$ potentials are chosen with respect to this criterion, not to the phases, the $^{1}S_{0}$ $np$ and $pp$ potentials excepted~\cite{NishonovValidation}; the $\alpha N$ $P$- and $D$-wave potentials have support points at positive energy only and a correspondingly larger $\Xi$, given in Ref.~\cite{NishonovValidation}. Three original interactions are represented, listed with their waves and reduced masses in Table~\ref{tab:parents} of Appendix~\ref{app:tables}: the charge-dependent CD~Bonn nucleon--nucleon ($NN$) potential~\cite{Machleidt2001}, the Kanada--Kaneko--Nagata--Nomoto (KKNN) $\alpha N$ potential~\cite{KKNN1979} and the Buck--Friedrich--Wheatley (BFW) $\alpha\alpha$ potential~\cite{BFW1977}. The $\alpha N$ $^{2}S_{1/2}$ wave contains the Pauli-forbidden $0s$ state, which is placed in the forbidden block, and the $\alpha p$ Coulomb force enters through the Fock block. The Coulomb part of the BFW potential is the point Coulomb force folded with the $\alpha$ charge distribution. The separable potential, in the $L=0$, $2$ and $4$ waves, is a fit of the original potential with the difference between its folded and the point Coulomb force absorbed into the short-range part, so that the separable potential plus the point Coulomb force of the Fock block reproduces the original potential with its folded Coulomb force.  Its Pauli-forbidden $0s$, $1s$ and $0d$ states form the forbidden block. The forbidden states are the Coulomb-dressed eigenstates of the extended separable problem (Sec.~\ref{sec:exact}); the quantities of the separable potentials with the Coulomb interaction included are collected in Appendix~\ref{app:tables}.

\subsection{The extended separable problem}
\label{sec:kernel}

With separable input every pair channel is a set of form factors and a coupling matrix, and the Fock modes are added to it as further terms; the same extended problem enters the three-body kernel of Ref.~\cite{FockII} unchanged.  The extended basis of a charged channel is
\begin{equation}
\bm{g} \;=\; \{\underbrace{\chi_{1},\ldots,\chi_{N_{r}}}_{\text{nuclear}},\;
\underbrace{\Phi_{0},\ldots,\Phi_{N_{C}-1}}_{\text{Fock}},\;
\underbrace{\phi_{1},\ldots,\phi_{N_{f}}}_{\text{forbidden}}\},
\label{eq:basisorder}
\end{equation}
with $\chi_{i}$, $i=1,\dots,N_{r}$, the nuclear form factors of Eq.~(\ref{eq:est}), $\Phi_{n}$, $n=0,\dots,N_{C}-1$, the Fock modes $\Phi_{nL}$ of Eq.~(\ref{eq:fockL}) in the partial wave $L$ of the channel, and $\phi_{f}$, $f=1,\dots,N_{f}$, the forbidden states~(\ref{eq:fbff}), and with the block-diagonal coupling matrix
\begin{equation}
\tilde{\bm\Lambda}=\mathrm{diag}\bigl(\bm\Lambda,\{\lambda_{nL}\},\lambda\,\mathbf{1}_{N_{f}}\bigr),
\label{eq:Lamext}
\end{equation}
whose three blocks are the fitted nuclear block, the diagonal Coulomb block with the analytic strengths~(\ref{eq:lamn}), (\ref{eq:lamnL}), and the forbidden block which effects the projection, $\lambda$ being the strength of the orthogonalizing pseudopotential, whose $\lambda\to\infty$ limit is taken below in closed form. No other change is required: the propagator integrals are computed for the $\Phi_{n}$ as for the nuclear form factors, and the pair $\tau$~matrix is the inverse of $[\tilde{\bm\Lambda}^{-1}-\tilde{\bm{\mathcal D}}(z)]$ over the extended basis. In a coupled charged channel the Fock modes are included with the lower orbital momentum of the pair, and no separate Coulomb modes are included for the higher component (for the $pp$ $^{3}P_{2}$--$^{3}F_{2}$ channel, $L=1$ only); the Coulomb block omitted in that component is small, the Coulomb force in the whole $pp$ $^{1}D_{2}$ wave being worth $5$~keV in $^{3}$He (Ref.~\cite{FockII}), and the $F$ component less. The price is an increase of the rank by $N_{C}\simeq20$ per charged channel; for bound states and other negative-energy quantities $N_{C}=20$ is enough: the $\alpha p$ $0s$ state is converged at $N_{C}=20$ (Table~\ref{tab:pauli}). The three-body equations with this input are the subject of Ref.~\cite{FockII}.

\subsection{The limit of validity: the behaviour of a finite truncation at positive energy}
\label{sec:boundary}

For a bound state the expansion converges rapidly. For a scattering phase relative to free waves it does not converge, as expected, since such a phase diverges for the unscreened Coulomb interaction: the representation is an identity for the potential, not an approximation scheme, and its convergence follows the analytic structure of the problem.  The limit is stated quantitatively because the extension to positive energy is constructed accordingly.

Every on-shell phase of this paper relative to free waves is the standing-wave phase of a finite-rank separable potential over its basis $\bm g$,
\begin{equation}
\tan\delta_{L}(k)=-\pi\mu k\;\bm g(k)^{T}\bigl[\bm\Lambda^{-1}-\bm{\mathcal D}^{P}(E_{k})\bigr]^{-1}\bm g(k),
\qquad E_{k}=\frac{k^{2}}{2\mu},
\label{eq:freephase}
\end{equation}
with $\bm{\mathcal D}^{P}$ the principal-value part of the propagator matrix of Eq.~(\ref{eq:tau}), the phase being defined modulo $180^{\circ}$; for the pure Coulomb block the basis is the $N_{C}$ Fock modes with the strengths~(\ref{eq:lamn}), for ``nuclear plus truncated Coulomb'' the extended basis~(\ref{eq:basisorder}) with the coupling matrix~(\ref{eq:Lamext}), the forbidden block eliminated by the Schur complement where the channel has forbidden states (Eq.~(\ref{eq:veff}) below at $\eta=0$).  Table~\ref{tab:boundary} shows two things.  First, the on-shell phase of the pure Coulomb block alone at $E_{\rm cm}=0.1$~MeV in the $\alpha p$ channel does not converge with the rank; this is the behaviour which the logarithmic on-shell singularity produces in any finite expansion in smooth functions.  Second, at rank $N_{C}$ the sum~(\ref{eq:fock}) is a screened Coulomb potential whose effective screening radius $R_{N}$ (Eq.~(\ref{eq:RN})) grows with $N_{C}$, and the phase of ``nuclear plus truncated Coulomb'' relative to free waves varies with $N_{C}$ as a screened Coulomb phase does, by $-\eta\ln(2kR_{N})$, where $\eta=Z_{1}Z_{2}e^{2}\mu/k$ is the Sommerfeld parameter.  In the $\alpha p$ $^{2}S_{1/2}$ channel at $E_{\rm cm}=20$~MeV it changes by $10^{\circ}$ between $N_{C}=10$ and $160$, away from the exact Coulomb-modified phase (Table~\ref{tab:pairs}), with a slope of order $-\eta$ per unit of $\ln N_{C}$, which identifies the $N_{C}$ dependence as a screening phase (Fig.~\ref{fig:screenlaw} shows the law itself).

\begin{table}[htbp]
\caption{The truncated Fock block at positive energy in the $\alpha p$ $^{2}S_{1/2}$ channel ($\beta_{C}=0.8$~fm$^{-1}$, with the strict projection) as a function of the rank $N_{C}$: the on-shell phase (degrees) of the pure Coulomb block alone at $E_{\rm cm}=0.1$~MeV, and the phase of nuclear plus truncated Coulomb relative to free waves at $E_{\rm cm}=20$~MeV.}
\label{tab:boundary}
\begin{center}
\begin{tabular}{lcccccc}
\toprule
$N_{C}$ & $10$ & $20$ & $40$ & $60$ & $80$ & $160$ \\
\midrule
pure Coulomb block, $0.1$~MeV        & $-44.36$ & $81.71$ & $41.37$ & $16.09$ & $-0.86$ & $-37.64$ \\
nuclear $+$ truncated Coulomb, $20$~MeV & $68.19$ & $65.92$ & $63.32$ & $61.93$ & $60.87$ & $58.38$ \\
\bottomrule
\end{tabular}
\end{center}
\end{table}

A finite rank therefore represents a screening, and the distinction between the negative- and positive-energy regimes is not sharp.

This is exploited in Sec.~\ref{sec:abovebreak}: the screening is removed in the usual way, by subtraction of the pure Coulomb phase of the same truncation, and the renormalized phase converges to the exact one.

\section{The eigenstate condition with the Coulomb force}
\label{sec:exact}

The Feshbach--Schur elimination of a Pauli-forbidden state, a projection in the sense of Feshbach~\cite{Feshbach1958,Feshbach1962}, is the $\lambda\to\infty$ limit of the orthogonalizing pseudopotential $\lambda\sum_{f}\ket{\phi_{f}}\bra{\phi_{f}}$~\cite{Kukulin1974}, which shifts the forbidden states $\phi_{f}$ to high energy at large coupling $\lambda$, and the limit holds for any projected vector~\cite{Nishonov2026a}; Fujiwara, Kohno and Suzuki~\cite{Fujiwara2004OCM} proved the same equivalence with the orthogonality-condition model also for the oscillator function, which is not an eigenstate of the pair potential. In coordinate space, for one $S$-wave forbidden state $\phi_f$ of a pair with $H=H_{0}+V$, the exact projected equation for the allowed component $w_a$ contains the nonlocal kernel
\begin{equation}
\mathcal K_{\rm exact}[w_a](r)
=\phi_f^{*}(r)\int dr'\,\bigl[(H-E)\phi_f\bigr](r')\,w_a(r'),
\label{eq:kexact}
\end{equation}
If $\phi_f$ is an eigenstate of the pair Hamiltonian actually used in the calculation,
\begin{equation}
H\phi_f=E_{f}\phi_f,
\label{eq:eigcond}
\end{equation}
the kernel is $(E_{f}-E)\phi_f^{*}(r)\langle\phi_f|w_a\rangle$ and vanishes on the allowed component, which is orthogonal to $\phi_f$: the projection removes the forbidden bound state and leaves the scattering states of $H$ unchanged~\cite{Nishonov2026a}. For a model $\phi_f$ the kernel does not vanish. In the separable formulation the Schur complement over the forbidden block of the extended coupling matrix~(\ref{eq:Lamext}), called below the strict projection (the Feshbach--Schur projection of Ref.~\cite{Nishonov2026a}, as distinct from the finite-strength pseudopotential), removes exactly the forbidden bound state and leaves the scattering states of the allowed block unchanged only when $\phi_f$ is a bound state of the extended separable problem from which it is removed. For any other vector the elimination is still exact as an operation, but it removes a different state: part of the true forbidden state remains, and the continuum is modified. The scattering states of a Hamiltonian are orthogonal to its own bound states, so the projection of an eigenstate cannot change them, while any other vector contains an admixture of scattering states, which the projection removes.  When the Coulomb force is added, the Hamiltonian changes and the orthogonal bound state changes accordingly and becomes the dressed state; the projection of the undressed state then modifies the continuum by the Coulomb-polarization admixture of that state; both effects are shown in the two-body comparisons below.

With the Coulomb block present, the Hamiltonian of Eq.~(\ref{eq:eigcond}) contains the Coulomb force, so the projected states must be the Coulomb-\emph{dressed} forbidden states, the bound states of nuclear plus Coulomb, not those of the nuclear potential alone and not the oscillator functions from which the number of forbidden states was derived. The dressed state itself is a well-known quantity, the bound state of the two-cluster Hamiltonian with its Coulomb force, which any coordinate-space treatment of the pair can produce, and the bound-state forbidden states of the $3\alpha$ calculations of Ref.~\cite{Fujiwara2004OCM} are of this kind.  Inside a momentum-space Faddeev calculation with separable input, however, this state was not available. With a screened Coulomb force the pair Hamiltonian in the kernel is a different operator at every screening radius, so its forbidden eigenstate is a screening-dependent quantity and its strict projection is a limit rather than a construction; with Coulomb-distorted form factors the forbidden state is not an eigenstate of any operator in the kernel. In the present separable potential the unscreened Coulomb force is inside the same separable problem, so the dressed states are eigenstates of the extended separable problem, obtained directly from it, and the Schur complement is taken over the last block of Eq.~(\ref{eq:basisorder}); by the eigenstate condition this state is the required one. Table~\ref{tab:dressed} gives the forbidden states of the separable potentials used here without and with the Coulomb force, each with the reduced mass of its channel; the dressed states are converged at $N_{C}=20$ to the digits quoted. They are used for the projection throughout this paper, except in the tests of the eigenstate condition in Sec.~\ref{sec:prescription}.

\begin{table}[tbp]
\caption{Pauli-forbidden states of the separable potentials (MeV) without the Coulomb force and Coulomb-dressed, and the Coulomb displacement; for $\alpha N$ the two columns are the $\alpha n$ and the $\alpha p$ state. The dressed states are at $N_{C}=20$, with $\beta_{C}=0.8$~fm$^{-1}$ for $\alpha p$ and $1.0$~fm$^{-1}$ for $\alpha\alpha$.}
\label{tab:dressed}
\begin{center}
\begin{tabular}{llccc}
\toprule
pair & state & Coulomb off & Coulomb on & displacement \\
\midrule
$\alpha N$ (KKNN)     & $0s$ & $-12.263$ & $-10.001$ & $2.26$\textsuperscript{a} \\
$\alpha\alpha$ (BFW)  & $0s$ & $-80.294$ & $-72.53$  & $7.76$ \\
                      & $1s$ & $-31.133$ & $-25.87$  & $5.26$ \\
                      & $0d$ & $-25.871$ & $-22.29$  & $3.58$ \\
\bottomrule
\end{tabular}
\par\noindent{\footnotesize \textsuperscript{a}~Difference of the $\alpha n$ state at the $\alpha n$ reduced mass and the $\alpha p$ state at the $\alpha p$ reduced mass; at the $\alpha p$ mass alone the displacement is $2.243$~MeV (Table~\ref{tab:pauli}).}
\end{center}
\end{table}

\section{Comparison with the other methods}
\label{sec:routes}

Three methods for the Coulomb force in a kernel with separable input are considered, and each is first written in the separable formalism, so that the benchmarks compare well-defined quantities. The screening method is the reference for the benchmarks: the exact two-body solution with a screened Coulomb force, Yukawa screening in the first benchmark and $\exp[-(r/R)^{4}]$ in the others, converged in $R$ to the accuracy stated at each benchmark. The benchmarks compare the distortion scheme and the Fock scheme with that reference at the two-body level, where the exact Lippmann--Schwinger solutions exist; a further benchmark compares the Fock representation with the screening method on the Coulomb-modified phase itself. The three-body benchmarks, on quantities to which no two-body construction can be tuned, are in Ref.~\cite{FockII}.

\subsection{The three methods in the separable formalism}
\label{sec:threemethods}

The first method is the screening with renormalization~\cite{Alt1978,Alt2002}. The point Coulomb force is replaced by the screened one,
\begin{equation*}
V_{C}^{R}(r)=Z_{1}Z_{2}e^{2}\,\frac{e^{-(r/R)^{s}}}{r},
\end{equation*}
with the screening radius $R$ and the exponent $s$ ($s=1$ is the Yukawa screening, $s=4$ the screening of Ref.~\cite{Deltuva2005} used below), whose partial-wave kernel
\begin{equation*}
V^{R}_{C,L}(p,p')=\frac{2}{\pi}\int_{0}^{\infty}r^{2}dr\,j_{L}(pr)\,V_{C}^{R}(r)\,j_{L}(p'r)
\end{equation*}
is regular at $p=p'$: the logarithm of the point kernel, Eq.~(\ref{eq:chebylog}) of Appendix~\ref{app:derivation}, is cut off at $|p-p'|\simeq1/R$. The screened Coulomb $t$~matrix is the solution of the Lippmann--Schwinger equation with this non-separable kernel,
\begin{equation*}
t^{R}_{C}(z)=V^{R}_{C}+V^{R}_{C}\,G_{0}(z)\,t^{R}_{C}(z),
\end{equation*}
a matrix equation on the momentum mesh, and the pair interaction $V^{R}_{C}+V_{s}$, with the separable nuclear part $V_{s}$ of Eq.~(\ref{eq:est}), has the two-potential $t$~matrix
\begin{equation}
t^{R}(z)=t^{R}_{C}(z)+\bigl[1+t^{R}_{C}(z)G_{0}(z)\bigr]
\sum_{ij}\ket{\chi_{i}}\bigl[\bm\Lambda^{-1}-\bm{\mathcal D}^{R}_{C}(z)\bigr]^{-1}_{ij}\bra{\chi_{j}}
\bigl[1+G_{0}(z)t^{R}_{C}(z)\bigr],
\label{eq:tscreen}
\end{equation}
with the propagator matrix of the screened Coulomb resolvent,
\begin{equation}
G^{R}_{C}(z)=(z-H_{0}-V^{R}_{C})^{-1}=G_{0}(z)+G_{0}(z)\,t^{R}_{C}(z)\,G_{0}(z),
\label{eq:GR}
\end{equation}
\begin{equation}
\bigl[\bm{\mathcal D}^{R}_{C}(z)\bigr]_{ij}=\braket{\chi_{i}|G^{R}_{C}(z)|\chi_{j}}
=\mathcal D_{ij}(z)+\braket{\chi_{i}|G_{0}(z)\,t^{R}_{C}(z)\,G_{0}(z)|\chi_{j}}.
\label{eq:DR}
\end{equation}
Equation~(\ref{eq:tscreen}) is identical to the direct solution of the Lippmann--Schwinger equation with the kernel $V^{R}_{C,L}+V_{s}$ on the mesh, and it shows where the screening enters a separable calculation: the separable form of the nuclear part is kept, but the pure Coulomb term $t^{R}_{C}$ and the two factors which dress the separable term are non-separable, and the propagator matrix is no longer the free one. On shell the phase of $t^{R}$ is $\delta^{R}_{L}(k)$ and that of $t^{R}_{C}$ is $\delta^{R}_{C,L}(k)$; both depend on $R$ and diverge logarithmically as $R\to\infty$, while their difference converges,
\begin{equation}
\delta_{SC,L}(k)=\lim_{R\to\infty}\bigl[\delta^{R}_{L}(k)-\delta^{R}_{C,L}(k)\bigr],
\qquad
\delta^{R}_{C,L}(k)=\sigma_{L}(\eta)+\phi_{R}(k)+o(1),
\label{eq:screenren}
\end{equation}
where $\sigma_{L}=\arg\Gamma(L+1+i\eta)$ is the pure Coulomb phase, $o(1)$ stands for the terms which vanish as $R\to\infty$, and
\begin{equation*}
\phi_{R}(k)=-\eta\bigl[\ln(2kR)-\gamma_{\scriptscriptstyle\mathrm{E}}/s\bigr]
\end{equation*}
is the renormalization phase of Refs.~\cite{Alt1978,Taylor1974}, derived in Appendix~\ref{app:derivation}, Eq.~(\ref{eq:aszphase}); the difference is the Coulomb-modified nuclear phase, the phase relative to Coulomb waves. In the benchmarks below the computed $\delta^{R}_{C,L}$ at the same radius is subtracted, not its asymptotic form. In the three-body equations the same $t^{R}$ enters the kernel as a non-separable term, the screened transition amplitudes are renormalized by phase factors of the same kind for the two-cluster states, and the limit $R\to\infty$ is taken numerically from a sequence of radii~\cite{Alt1978,Deltuva2005}; every quantity depends on $R$ until the limit is taken.

The second method is that of the Coulomb-distorted form factors. The nuclear form factors are dressed with the M{\o}ller wave operator of the screened Coulomb potential, the operator which maps free states onto its scattering states,
\begin{equation}
g^{R}_{i}(z)=\bigl[1+t^{R}_{C}(z)G_{0}(z)\bigr]\chi_{i},
\qquad
g^{R}_{i}(p;z)=\chi_{i}(p)+\int_{0}^{\infty}dp'\,p'^{2}\,
\frac{t^{R}_{C}(p,p';z)\,\chi_{i}(p')}{z-p'^{2}/2\mu},
\label{eq:dressedff}
\end{equation}
which is the factor multiplying the separable term in Eq.~(\ref{eq:tscreen}); the first-order dressing is Eq.~(\ref{eq:dressedff}) with $t^{R}_{C}$ replaced by $V^{R}_{C}$, the dressing is iterated to all orders unless first order is stated, and the screening is $\exp[-(r/R)^{4}]$ as in Ref.~\cite{Deltuva2005}. The Coulomb--nuclear interference is then contained in the form factors, and the calculation keeps the separable form,
\begin{equation*}
t_{\rm dist}(z)=\sum_{ij}\ket{g^{R}_{i}(z)}\bigl[\bm\Lambda^{-1}-\bm{\mathcal D}^{R}_{\rm dist}(z)\bigr]^{-1}_{ij}\bra{g^{R}_{j}(z)},
\end{equation*}
\begin{equation*}
\bigl[\bm{\mathcal D}^{R}_{\rm dist}(z)\bigr]_{ij}=\braket{g^{R}_{i}(z)|G_{0}(z)|g^{R}_{j}(z)},
\end{equation*}
with the coupling matrix of the nuclear fit and the free resolvent in the propagator integrals: the Coulomb force enters only through the form factors. This scheme, Coulomb-dressed form factors with the free resolvent in the propagator integrals, is the screened variant of the form-factor construction of the Coulomb-basis formulation of Refs.~\cite{Mukhamedzhanov2012,Hlophe2014,Eremenko2015,Hlophe2017}; that formulation propagates with the Coulomb Green's function between the distorted form factors, the two-potential form of Eq.~(\ref{eq:tscreen}), and neither it nor its unscreened regularized folding is tested here. Compared with the exact Eq.~(\ref{eq:tscreen}) two terms are absent. The pure Coulomb term $t^{R}_{C}$, which has no separable form in this scheme, is absent, and the propagator matrix differs from Eq.~(\ref{eq:DR}): by Eq.~(\ref{eq:GR}),
\begin{equation}
\bigl[\bm{\mathcal D}^{R}_{\rm dist}-\bm{\mathcal D}^{R}_{C}\bigr]_{ij}
=\braket{\chi_{i}|G_{0}t^{R}_{C}G_{0}|\chi_{j}}
+\braket{\chi_{i}|G_{0}t^{R}_{C}G_{0}t^{R}_{C}G_{0}|\chi_{j}},
\label{eq:distdiff}
\end{equation}
so that the term of first order in the Coulomb force is counted twice, once from each dressed form factor. A bound state of the scheme, a root of
\begin{equation*}
\det\bigl[\bm\Lambda^{-1}-\bm{\mathcal D}^{R}_{\rm dist}(E)\bigr]=0,
\end{equation*}
is therefore displaced by twice the first-order Coulomb shift, and the scheme is not the two-potential formula with the Coulomb propagator matrix, which would be exact. The scheme is defined at the negative pair energies which the kernel samples; it defines no on-shell phase.

The third method is the Fock representation, Eqs.~(\ref{eq:basisorder}) and~(\ref{eq:Lamext}): the Coulomb block is a further set of separable terms with analytic strengths, the propagator matrix is the free one over the extended basis, and no screening is involved; the truncation error at negative energy falls with $N_{C}$. On shell, where a finite rank is the screening of Eq.~(\ref{eq:RN}), the counterpart of Eq.~(\ref{eq:screenren}) is the subtraction of the pure phase of the truncated block, Eq.~(\ref{eq:renphase}), with the rank $N_{C}$ in the place of $R$.

The constants are $\hbar c = 197.327$~MeV\,fm and $e^{2}=\alpha\hbar c=1.440$~MeV\,fm ($\alpha=1/137.036$ the fine-structure constant; $e^{2}$ is used for it throughout, since $\alpha$ denotes the $\alpha$ particle), $m_{p}=938.272$ and $m_{n}=939.565$~MeV, reduced masses formed from these values without rounding; the cluster masses are $m_{\alpha}=4m_{N}$ with $m_{N}=939.08$~MeV, and the cluster reduced masses are those of Table~\ref{tab:parents}.

\subsection{Benchmark I: the off-shell $t$~matrix}
\begin{table}[b]
\caption{Relative deviation of the off-shell $t(p,p';E)$ of each Coulomb scheme from the exact solution, $\alpha p$ $^{2}P_{3/2}$: mean and maximum over the fifteen test points, against the Yukawa-screened reference at each of its two screening radii; the last row is the change of the reference itself between the radii. The Fock modes are at $N_{C}=20$, $\beta_{C}=0.8$~fm$^{-1}$.}
\label{tab:offshell}
\begin{center}
\begin{tabular}{lcccc}
\toprule
 & \multicolumn{2}{c}{reference at $R=20$~fm} & \multicolumn{2}{c}{reference at $R=100$~fm} \\
\cmidrule(lr){2-3}\cmidrule(lr){4-5}
scheme & mean $|$dev$|$ & max $|$dev$|$ & mean $|$dev$|$ & max $|$dev$|$ \\
\midrule
Fock modes, no distortion        & $0.5\%$  & $2.3\%$  & $0.7\%$  & $2.4\%$ \\
distortion, no Fock              & $12.4\%$ & $22.2\%$ & $12.3\%$ & $22.1\%$ \\
distortion $+$ Fock (both)       & $30.0\%$ & $44.7\%$ & $29.9\%$ & $44.7\%$ \\
\midrule
reference, $R=20$ vs $100$~fm & $0.21\%$ & $0.79\%$ & --- & --- \\
\bottomrule
\end{tabular}
\end{center}
\end{table}

\label{sec:offshell}

The three-body kernel samples the pair interaction at negative pair energies $z-q^{2}/2\bar{\mu}$, where $z$ is the three-body energy and $\bar{\mu}$ the reduced mass of the spectator with respect to the pair; a bound-state kernel never samples it on shell. The first benchmark is therefore the off-shell $t(p,p';E)$ of the $\alpha p$ $^{2}P_{3/2}$ pair.  The reference is the direct solution of the Lippmann--Schwinger equation at $E<0$ with the KKNN force plus the partial-wave Coulomb force, screened in the Yukawa form $e^{-r/R}/r$ for this reference; the Fock rows are at $N_{C}=20$, $\beta_{C}=0.8$~fm$^{-1}$. Each scheme is evaluated in the form in which it enters the Faddeev kernel, at fifteen test points, three energies and five momentum pairs: $E=-1,-3,-5$~MeV and the five off-diagonal momentum pairs $(p,p')=(0.3,0.6)$, $(0.6,1.0)$, $(1.0,1.5)$, $(1.5,2.0)$ and $(0.3,1.5)$~fm$^{-1}$. The reference changes by less than a percent between $R=20$ and $100$~fm (last row of Table~\ref{tab:offshell}).

The Fock representation reproduces the exact off-shell $t$~matrix at the percent level (Table~\ref{tab:offshell}). The distortion scheme underestimates the Coulomb repulsion by about ten per cent: the dressing includes the Coulomb--nuclear interference but omits the pure Coulomb term $t^{R}_{C}$ of Eq.~(\ref{eq:tscreen}), which has no separable form in that scheme. With both schemes applied together in the same channel the deviation doubles: the Fock block provides the pure Coulomb term which is absent in the distortion scheme, but the dressing acts on every form factor of the basis, the Fock modes included, and the Coulomb force is counted twice.

\subsection{Benchmark II: the $\alpha p$ Pauli-state displacement}
\label{sec:pauli}

The Coulomb displacement of the $\alpha p$ $0s$ forbidden state is a bound-state quantity computable identically in every scheme, and it determines the dressed projection. Its exact value (Table~\ref{tab:pauli}) is obtained twice, by coordinate-space integration of the KKNN potential plus the point Coulomb force and by a separable potential of the full nuclear-plus-Coulomb interaction fitted at support points at negative energy with the Coulomb force screened as $\exp[-(r/R)^{4}]$ at $R=100$~fm, where the screening is negligible at the radius of the orbit; the two agree.

\begin{table}[htbp]
\caption{The $\alpha p$ $0s$ Pauli state in each scheme (MeV) and its Coulomb displacement from the nuclear-only $\alpha p$ state at the $\alpha p$ reduced mass ($-12.2439$~MeV; the $\alpha n$ state of Table~\ref{tab:dressed} differs by the reduced mass); $R$ is the screening radius of the distortion scheme.}
\label{tab:pauli}
\begin{center}
\begin{tabular}{lcc}
\toprule
scheme & $E_{0s}$ (MeV) & displacement (MeV) \\
\midrule
Fock, $N_{C}=10$            & $-10.0011$ & $+2.2428$ \\
Fock, $N_{C}=20$            & $-10.0009$ & $+2.2430$ \\
Fock, $N_{C}=30$            & $-10.0009$ & $+2.2430$ \\
exact, coordinate space     & $-9.9982$  & $+2.2457$ \\
exact, momentum space       & $-9.9984$  & $+2.2455$ \\
distortion, first order, $R=20$~fm  & $-7.072$  & $+5.17$  \\
distortion, all orders, $R=20$~fm   & $-7.612$  & $+4.63$  \\
distortion, $R=40$~fm       & $-7.619$  & $+4.62$  \\
distortion, $R=80$~fm       & $-7.691$  & $+4.55$  \\
\bottomrule
\end{tabular}
\end{center}
\end{table}

The Fock scheme reproduces the displacement to a few keV; the residual is the finite-rank error of the nuclear separable potential, not of the Coulomb block. The distortion scheme overestimates the displacement by about a factor of two, changes by half an MeV between the first-order and the all-orders dressing, and its sequence in $R$ does not approach the exact value. Its error here overestimates the Coulomb repulsion, the opposite sign to its off-shell error above; it is therefore not a normalization error that a single factor could correct. The factor of two is the one of Eq.~(\ref{eq:distdiff}): the propagator matrix of the scheme counts the first-order Coulomb term twice, and the bound-state energy is shifted accordingly.

One systematic error is common to all schemes. All use the point-charge $\alpha p$ Coulomb force, while folding with the charge distribution of the $\alpha$ particle gives $2e^{2}\,\mathrm{erf}(\beta_{\alpha}r)/r$, with $\beta_{\alpha}\simeq0.73$~fm$^{-1}$ for the $\alpha$ charge radius of $1.68$~fm; the folding, weaker than the point force at every radius, would lower the dressed $0s$ state; its size is not computed here. No relative comparison here depends on it; the absolute Coulomb shift of $^{6}$Li in Ref.~\cite{FockII} is subject to it.

\subsection{Benchmarks III and IV: the two methods on shell}
\label{sec:onshell}\label{sec:pairs}

The on-shell phases, unlike the two benchmarks above, are observables, and two on-shell comparisons follow. The first compares two constructions, neither of which is the Fock method; it tests the alternative of fitting the Coulomb force inside the separable potential.  The Coulomb-modified $\alpha p$ nuclear phase $\delta_{SC,L}$ is computed in two ways with the same reduced mass (Table~\ref{tab:onshell}). The first is the renormalized $\delta_{SC}$ of the local KKNN potential: the point Coulomb force is screened as $\exp[-(r/R)^{4}]$ at $R=100$~fm and the phase of the screened Coulomb force alone is subtracted, and this is the reference. The second is the negative-energy-support separable potential used for the Pauli state, the separable fit of the full nuclear-plus-Coulomb $t$~matrix at negative energies, evaluated on shell without subtraction of a Coulomb reference.

\begin{table}[htbp]
\caption{Coulomb-modified $\alpha p$ nuclear phases: largest disagreement between the renormalized original-potential reference and the negative-energy-support separable potential evaluated on shell, the energy at which it occurs, and the largest disagreement below $E_{\rm cm}=1$~MeV.}
\label{tab:onshell}
\begin{center}
\begin{tabular}{lrlr}
\toprule
wave & $\max|\Delta\delta_{SC}|$ & at & $E_{\rm cm}\le1$~MeV \\
\midrule
$^{2}P_{3/2}$ & $10.46^{\circ}$ & $3.0$~MeV  & $\le6.10^{\circ}$ \\
$^{2}P_{1/2}$ & $12.51^{\circ}$ & $6.0$      & $\le0.66^{\circ}$ \\
$^{2}D_{5/2}$ & $5.25^{\circ}$  & $18.0$     & $<0.01^{\circ}$ \\
\bottomrule
\end{tabular}
\end{center}
\end{table}

The $^{2}S_{1/2}$ wave is treated in the projected scattering calculation below, since unprojected its phase is defined only up to the multiple of $\pi$ which Levinson's theorem assigns to the forbidden bound state; the $^{2}D_{3/2}$ wave is omitted. The two constructions agree where the phases are small, in $^{2}D_{5/2}$ below $1$~MeV, and disagree by five to thirteen degrees where the phases become large. At $R=200$~fm the maxima change by less than a degree and the bounds below $1$~MeV by much less; this is the remaining dependence of the renormalized reference on the screening. This pattern is characteristic of a separable potential which coincides with the exact $t$~matrix at its support points at negative energy and deviates from it on shell. The reason is general and underlies the extension to positive energy: a finite-rank $t$~matrix built from smooth form factors has a phase which goes to zero at threshold as $k^{2L+1}$, while the pure Coulomb phase $\sigma_{L}$ diverges logarithmically there. A separable potential with the Coulomb force inside its fit reproduces the Coulomb effects off shell: its dressed Pauli state is correct (Table~\ref{tab:pauli}). It cannot reproduce them on shell, where the singularity lies. Agreement at negative energy is therefore no evidence that the two constructions agree on shell, because the kernel samples the pair interaction below threshold only.

The second on-shell comparison, of the Fock representation with the screening method, is made for the Coulomb-modified nuclear phase $\delta_{SC}$ of the three charged pairs of this paper, $pp$ ($^{1}S_{0}$), $\alpha p$ ($^{2}S_{1/2}$, with the strict projection) and $\alpha\alpha$ ($L=0$, with the strict projection), against the exact construction of the Coulomb-modified phase on the same separable potentials, the two-potential formula with the closed-form Coulomb Green's function given in Sec.~\ref{sec:twobodyscat}. The screening method is the renormalized phase of the pair with the Coulomb force screened as $\exp[-(r/R)^{4}]$ at $R=20$, $40$ and $80$~fm; the Fock representation gives the renormalized phase at rank $N_{C}=20$, $80$ and $320$. Table~\ref{tab:pairs} gives the deviations from the exact phase and Fig.~\ref{fig:phases} the phases themselves. The distortion method defines no on-shell phase (Sec.~\ref{sec:threemethods}) and is not included.

\begin{table}[htbp]
\caption{Deviation of the Coulomb-modified nuclear phase from the exact construction (degrees): the screening method with $\exp[-(r/R)^{4}]$ at $R=20/40/80$~fm, and the Fock representation at $N_{C}=20/80/320$; exact phases in the last column. Deviations of a thousandth of a degree or less are zero within the numerical precision.}
\label{tab:pairs}
\begin{center}
\footnotesize
\begin{tabular}{lc|rrr|rrr|r}
\toprule
pair & $E_{\rm cm}$ (MeV) & \multicolumn{3}{c|}{screening, $R=20/40/80$} & \multicolumn{3}{c|}{Fock, $N_{C}=20/80/320$} & exact \\
\midrule
$pp$ $^{1}S_{0}$ & $0.5$  & $-1.229$ & $+0.019$ & $-0.001$ & $-0.729$ & $-0.280$ & $+0.052$ & $32.803$ \\
                 & $2.5$  & $-0.013$ & $+0.001$ & $-0.000$ & $+0.094$ & $-0.063$ & $+0.017$ & $54.870$ \\
                 & $12.5$ & $+0.001$ & $+0.001$ & $+0.000$ & $-0.049$ & $+0.007$ & $-0.004$ & $48.624$ \\
                 & $25$   & $+0.001$ & $+0.001$ & $+0.001$ & $-0.014$ & $+0.009$ & $+0.001$ & $38.799$ \\
\midrule
$\alpha p$ $^{2}S_{1/2}$ & $0.5$ & $+0.396$ & $+0.009$ & $+0.001$ & $+0.473$ & $+0.147$ & $+0.031$ & $-8.047$ \\
                     & $2$   & $+0.016$ & $+0.001$ & $+0.000$ & $-0.109$ & $+0.136$ & $-0.007$ & $-29.721$ \\
                     & $5$   & $+0.001$ & $+0.000$ & $+0.000$ & $-0.330$ & $-0.052$ & $-0.025$ & $-53.354$ \\
                     & $20$  & $-0.000$ & $-0.001$ & $-0.001$ & $-0.170$ & $+0.043$ & $+0.008$ & $81.352$ \\
\midrule
$\alpha\alpha$ $L=0$ & $0.5$ & $-1.641$ & $-0.092$ & $-0.007$ & $+1.361$ & $+0.082$ & $-0.001$ & $-10.004$ \\
                     & $2$   & $-0.026$ & $-0.005$ & $-0.002$ & $+2.111$ & $-0.563$ & $-0.128$ & $-68.638$ \\
                     & $5$   & $-0.006$ & $-0.004$ & $-0.002$ & $-0.908$ & $-0.229$ & $+0.043$ & $51.958$ \\
                     & $20$  & $-0.001$ & $-0.007$ & $-0.004$ & $-0.439$ & $-0.118$ & $+0.025$ & $-52.772$ \\
\bottomrule
\end{tabular}
\end{center}
\end{table}

The two methods converge to the exact phase in different ways and at different rates. The screening method with $\exp[-(r/R)^{4}]$ agrees closely with the exact phase at $R\ge40$~fm in all three pairs at every energy; the deviation at $R=20$~fm is the screening error at the lowest energy. The Yukawa screening at the same radii deviates by one to eleven degrees at $0.5$~MeV and still by $0.7$--$3^{\circ}$ at $R=80$~fm, as observed in Ref.~\cite{Deltuva2005}. The Fock representation converges with a remainder of order $1/N_{C}$, oscillating in sign, and reaches the exact phase at $N_{C}=320$ in $pp$ and $\alpha p$ and at $N_{C}=640$ in $\alpha\alpha$ (Table~\ref{tab:pairs}, Fig.~\ref{fig:phases}), where the Coulomb force is strongest ($\eta=1.27$ at $0.5$~MeV). Both methods reproduce the exact Coulomb-modified phase. On shell the screening method is computationally cheaper. The Fock representation needs no screening function or radius; the $\alpha p$ Pauli state and the $^{8}$Be resonance of Table~\ref{tab:pairs2} are converged to the quoted digits for $N_{C}\ge80$.

The bound-state and threshold quantities of the same pairs show the same behaviour (Table~\ref{tab:pairs2}). The $\alpha p$ $0s$ Pauli state (Sec.~\ref{sec:pauli}) is reproduced by the $\exp[-(r/R)^{4}]$ screening at $R\ge40$~fm, where the Yukawa screening still gives a state tens of keV too deep. The Coulomb-modified $pp$ effective-range parameters are the quantities most sensitive to the long-range part of the interaction. Their exact values from the low-energy interval are those of the construction of Sec.~\ref{sec:twobodyscat}; the truncated methods cannot reach that interval (it would need $R>500$~fm or $N_{C}>2000$), so the comparison is made on a two-parameter fit over $E_{\rm cm}=0.05$--$1$~MeV, where the omitted shape term lowers $r_{C}$, against the exact values on the same interval. The screening method reproduces them at $R=320$~fm and the Fock representation approaches them as $1/N_{C}$. The $^{8}$Be ground-state resonance is reproduced by the Fock representation at $N_{C}\ge80$ and by the screening at $R=80$~fm; at $R=20$~fm the screening distorts the Coulomb barrier so strongly that the resonance disappears.

\begin{table}[htbp]
\caption{Bound-state and threshold quantities of the three charged pairs by method: the $\alpha p$ $0s$ Pauli state (MeV), the Coulomb-modified $pp$ effective-range parameters (fm), and the $^{8}$Be ground-state resonance (keV).  Screening with $\exp[-(r/R)^{4}]$ unless marked Yukawa; $R$ in fm. The exact $^{8}$Be value is the $\delta_{0}=90^{\circ}$ crossing of the Coulomb-modified phase of the separable potential (Appendix~\ref{app:tables}, Table~\ref{tab:coulon}), the Fock and screening entries the crossings of the respective phases. The pole of the two-pole form~(\ref{eq:fbff}) of the Fock-projected $\alpha p$ state lies at $-10.0005$~MeV.}
\label{tab:pairs2}
\begin{center}
\small
\begin{tabular}{llll}
\toprule
quantity & exact & Fock representation & screening \\
\midrule
$E_{0s}(\alpha p)$        & $-9.9982$ & $-10.0009$ ($N_{C}\ge20$) & $-9.9984$, $-9.9982$, $-9.9982$ \\
                          &            &                            & \quad($R=20$, $40$, $80$) \\
                          &            &                            & Yukawa: $-10.136$, $-10.069$, $-10.034$ \\
$a_{C}$, $r_{C}$ ($pp$)   & $-7.811$, $2.730$ & $-7.749$, $2.60$ ($N_{C}=320$) & $-7.806$, $2.721$ ($R=80$) \\
                          &            & $-7.801$, $2.71$ ($N_{C}=1280$) & $-7.811$, $2.729$ ($R=320$) \\
$E_{r}(^{8}\mathrm{Be},0^{+})$ & $105.94$ & $105.97$, $105.94$, $105.94$ & $105.18$, $105.96$ ($R=40$, $80$) \\
                          &            & \quad($N_{C}=20$, $80$, $320$) & \\
\bottomrule
\end{tabular}
\end{center}
\end{table}

\section{Deep versus oscillator forbidden states}
\label{sec:prescription}

\begin{figure}[htbp]
\begin{center}
\begin{tikzpicture}
\begin{groupplot}[group style={group size=2 by 1, horizontal sep=1.6cm},
  width=0.5\textwidth, height=5.6cm,
  tick label style={font=\scriptsize}, label style={font=\scriptsize},
  title style={font=\scriptsize},
  legend style={font=\scriptsize, draw=none, fill=none, inner sep=1pt, row sep=-3pt}, legend cell align=left]
\nextgroupplot[xlabel={$r$ (fm)}, ylabel={$u(r)$ (fm$^{-1/2}$)}, xmin=0, xmax=4, ymin=0, legend pos=north east]
\addplot[black] coordinates {(0,0.000002) (0.2,0.366767) (0.4,0.672000) (0.6,0.878424) (0.8,0.968893) (1,0.949617) (1.2,0.847537) (1.4,0.699364) (1.6,0.540545) (1.8,0.396759) (2,0.280591) (2.2,0.193585) (2.4,0.131056) (2.6,0.086647) (2.8,0.055022) (3,0.032628) (3.2,0.017319) (3.4,0.007655) (3.6,0.002390) (3.8,0.000274) (4,0.000097)}; \addlegendentry{Coulomb-dressed}
\addplot[black, dashed] coordinates {(0,0.000002) (0.2,0.417751) (0.4,0.736964) (0.6,0.930262) (0.8,0.992031) (1,0.940699) (1.2,0.812935) (1.4,0.650290) (1.6,0.488146) (1.8,0.348998) (2,0.241397) (2.2,0.163629) (2.4,0.109144) (2.6,0.070964) (2.8,0.043883) (3,0.024758) (3.2,0.011840) (3.4,0.003965) (3.6,0.000040) (3.8,-0.001102) (4,-0.000615)}; \addlegendentry{undressed}
\addplot[black, dotted, thick] coordinates {(0,0.000002) (0.2,0.320896) (0.4,0.599900) (0.6,0.804102) (0.8,0.915900) (1,0.935004) (1.2,0.876006) (1.4,0.762823) (1.6,0.622075) (1.8,0.477397) (2,0.345922) (2.2,0.237230) (2.4,0.154246) (2.6,0.095212) (2.8,0.055853) (3,0.031163) (3.2,0.016548) (3.4,0.008368) (3.6,0.004031) (3.8,0.001851) (4,0.000810)}; \addlegendentry{oscillator}
\nextgroupplot[xlabel={$p$ (fm$^{-1}$)}, ylabel={$\phi(p)$ (fm$^{3/2}$)}, ymode=log, xmin=0, xmax=4, ymin=0.0008, ymax=2, legend pos=south west]
\addplot[black] coordinates {(0,1.300003) (0.2,1.273073) (0.4,1.202084) (0.6,1.102379) (0.8,0.980795) (1,0.843773) (1.2,0.703467) (1.4,0.570657) (1.6,0.450436) (1.8,0.345404) (2,0.257913) (2.2,0.188806) (2.4,0.136374) (2.6,0.097289) (2.8,0.068185) (3,0.046523) (3.2,0.030606) (3.4,0.019243) (3.6,0.011466) (3.8,0.006412) (4,0.003317)}; \addlegendentry{Coulomb-dressed}
\addplot[black, dashed] coordinates {(0,1.211975) (0.2,1.188562) (0.4,1.127018) (0.6,1.040614) (0.8,0.934518) (1,0.813566) (1.2,0.688170) (1.4,0.567835) (1.6,0.457094) (1.8,0.358511) (2,0.274751) (2.2,0.207217) (2.4,0.154834) (2.6,0.114808) (2.8,0.084158) (3,0.060616) (3.2,0.042686) (3.4,0.029340) (3.6,0.019726) (3.8,0.013056) (4,0.008598)}; \addlegendentry{undressed}
\addplot[black, dotted, thick] coordinates {(0,1.375239) (0.2,1.351006) (0.4,1.280840) (0.6,1.171901) (0.8,1.034774) (1,0.881776) (1.2,0.725153) (1.4,0.575519) (1.6,0.440806) (1.8,0.325832) (2,0.232434) (2.2,0.160016) (2.4,0.106312) (2.6,0.068165) (2.8,0.042180) (3,0.025188) (3.2,0.014516) (3.4,0.008074) (3.6,0.004334) (3.8,0.002245) (4,0.001122)}; \addlegendentry{oscillator}
\end{groupplot}
\end{tikzpicture}
\end{center}
\caption{The three vectors of the $\alpha\alpha$ $0s$ forbidden state, normalized to unity: the Coulomb-dressed eigenstate of the extended separable problem, the undressed (Coulomb-subtracted) eigenstate, and the oscillator function of Table~\ref{tab:ho}; the radial function $u(r)=rR(r)$ (left) and the momentum-space form on a logarithmic scale (right).}
\label{fig:fbstates}
\end{figure}
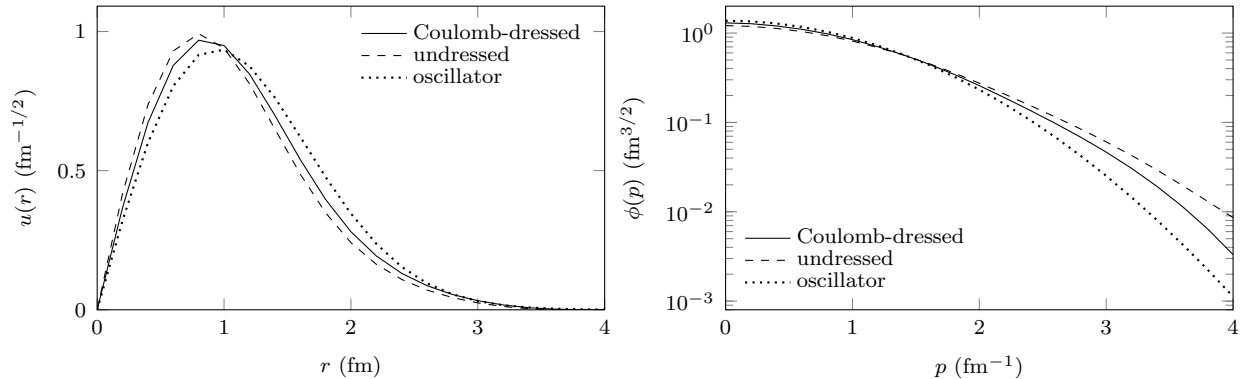

The eigenstate condition can be tested directly, because two prescriptions for the projected function are in use. The \emph{deep-eigenstate} prescription, used for every result of this paper outside the comparison, takes the forbidden state as the bound state of the separable potential itself, Coulomb-dressed where the Coulomb force is on; it satisfies the condition by construction. The \emph{harmonic-oscillator} prescription of the orthogonality-condition model takes the microscopic oscillator functions, with the widths of Table~\ref{tab:ho} (Appendix~\ref{app:ho}); these are close to the eigenstates of the interactions used here but do not coincide with them, so the elimination removes a slightly different state. Figure~\ref{fig:fbstates} shows the three vectors for the $\alpha\alpha$ $0s$ state, the dressed and undressed eigenstates of Table~\ref{tab:dressed} and the oscillator function. In coordinate space they nearly coincide, the overlap of Table~\ref{tab:ho} for this state being within half a percent of unity, the oscillator function the widest of the three. In momentum space, on a logarithmic scale, the oscillator function falls as a Gaussian where the eigenstates of the two-pole form~(\ref{eq:fbff}) fall as a power, and the Coulomb dressing shifts the eigenstate toward lower momenta. Beyond $4$~fm the two-pole forms have an oscillating tail below a thousandth of the peak, a property of the representation and not a node of the state.

That the two choices differ in the $3\alpha$ system is known: Fujiwara, Kohno and Suzuki~\cite{Fujiwara2004OCM} found $2$~MeV in the $^{12}$C ground state without the Coulomb force, the oscillator state being the less bound, and Tursunov and Mazumdar~\cite{Tursunov2022} a strong sensitivity of the $0^{+}$ and $2^{+}$ states to the elimination of the forbidden states, the deep-phase states being created by the critical eigenstates of the Pauli projector. Moriya, Horiuchi and Zhou~\cite{Moriya2023} construct a coordinate-space basis which is nearly orthogonal to the oscillator forbidden states by construction, so that the pseudopotential is not needed (negligible residual forbidden component, basis reduced from about $1000$ to about $100$ functions). That approach changes the basis; the present one changes the operator, at the price of one Schur complement per pair. The removal method itself, projection versus supersymmetric transformation, has been compared by Deltuva~\cite{Deltuva2026} for core-plus-two-nucleon systems, where the forbidden states are the deep eigenstates of the core--nucleon potential as here and an oscillator approximation is reported to make little difference; the $\alpha\alpha$ pair, which produces the effect found below, does not occur there.  The question asked here, which their construction does not raise, is whether the oscillator function is the appropriate state to remove. It is answered here at the two-body level, on the on-shell form of the test, with everything else kept fixed; the spectrum of the projected pair and the three-body form of the test, in $A=6$, $^{9}$Be and $^{12}$C, are in Ref.~\cite{FockII}.

\subsection{The two-cluster phase shifts}
\label{sec:conv2b}

For the two-cluster phases the test is the phase shift itself. The Coulomb-modified phase is computed with the strict Feshbach--Schur projection, once with the dressed eigenstate and once with the oscillator function removed from the same nuclear-plus-Coulomb interaction.  The difference is the on-shell effect of the violated condition, as a function of channel and energy, which the binding energies cannot show. Table~\ref{tab:conv2b} gives it for the channels which contain a forbidden state, on the separable potentials of this paper and on an otherwise identical separable potential in which only the forbidden states are replaced by the oscillator functions. For $\alpha N$ the $\alpha n$ channel, without the Coulomb interaction, is the reference; for $\alpha\alpha$ the removed eigenstate is the Coulomb-dressed state of the Coulomb-subtracted BFW separable potential with the point Coulomb force. Without the projection the two separable potentials give the same phase in every channel. Scattering states are orthogonal to a bound state, so the projection of a true eigenstate cannot change the phases; they can change only when the projected vector is not an eigenstate, and then by an amount which depends on the vector.

\begin{table}[htbp]
\caption{Coulomb-modified nuclear phases $\delta_{SC}$ (degrees) with the strict Feshbach--Schur projection of two vectors from the same interaction, its own forbidden eigenstate (eig., Coulomb-dressed where the Coulomb force is on) and the oscillator function (h.o.) at the width of the relative motion; diff.\ is h.o.\ minus eig. All phases are computed with the construction of Sec.~\ref{sec:twobodyscat}, the same as in Table~\ref{tab:phaseeq}; the $\alpha\alpha$ phases are those of the Coulomb-subtracted BFW separable potential with the point Coulomb force.}
\label{tab:conv2b}
\begin{center}
\begin{tabular}{c|rrr|rrr}
\toprule
$E_{\rm cm}$ & \multicolumn{3}{c|}{$\alpha p$ $^{2}S_{1/2}$} & \multicolumn{3}{c}{$\alpha n$ $^{2}S_{1/2}$ (Coulomb off)} \\
(MeV) & eig. & h.o. & diff. & eig. & h.o. & diff. \\
\midrule
 $0.1$ &  $-0.39$ &  $-0.33$ & $+0.07$ &  $-9.38$ &  $-8.48$ & $+0.89$ \\
 $0.5$ &  $-8.05$ &  $-6.83$ & $+1.22$ & $-20.83$ & $-18.90$ & $+1.93$ \\
 $1$   & $-16.83$ & $-14.51$ & $+2.32$ & $-29.22$ & $-26.61$ & $+2.61$ \\
 $2$   & $-29.72$ & $-26.08$ & $+3.64$ & $-40.68$ & $-37.29$ & $+3.39$ \\
 $5$   & $-53.35$ & $-48.25$ & $+5.10$ & $-61.55$ & $-57.38$ & $+4.17$ \\
 $10$  & $-75.53$ & $-70.48$ & $+5.05$ & $-81.39$ & $-77.56$ & $+3.84$ \\
 $20$  &  $81.35$ &  $84.36$ & $+3.01$ &  $77.81$ &  $79.76$ & $+1.94$ \\
\midrule
 & \multicolumn{3}{c|}{$\alpha\alpha$ $L=0$} & \multicolumn{3}{c}{$\alpha\alpha$ $L=2$} \\
\midrule
 $0.1$ &   $0.06$ &  $-0.00$ & $-0.06$ &   $0.00$ &   $0.00$ & $+0.00$ \\
 $0.5$ & $-10.00$ &  $-6.24$ & $+3.76$ &   $0.02$ &   $0.02$ & $+0.00$ \\
 $1$   & $-32.97$ & $-27.11$ & $+5.86$ &   $0.48$ &   $0.60$ & $+0.13$ \\
 $2$   & $-68.64$ & $-62.97$ & $+5.67$ &   $9.71$ &  $14.92$ & $+5.21$ \\
 $5$   &  $51.96$ &  $55.79$ & $+3.83$ & $-63.51$ & $-56.18$ & $+7.33$ \\
 $10$  &   $0.92$ &   $2.91$ & $+2.00$ & $-79.37$ & $-77.13$ & $+2.25$ \\
 $20$  & $-52.77$ & $-52.39$ & $+0.38$ &  $70.81$ &  $71.03$ & $+0.23$ \\
\bottomrule
\end{tabular}
\end{center}
\end{table}

In the $\alpha N$ channel the oscillator error is one to five degrees and is the same with the Coulomb interaction ($\alpha p$) as without it ($\alpha n$): the failure is that the oscillator function is not an eigenstate of the KKNN interaction, and the part specific to the Coulomb dressing is about one degree at most. In the $\alpha\alpha$ $L=0$ channel, where the removed state is the Coulomb-dressed eigenstate of the Coulomb problem, the oscillator error is up to six degrees between $0.5$ and $5$~MeV and has disappeared by $20$~MeV, and in $L=2$ it reaches seven degrees at $5$~MeV; the oscillator functions lack up to $2\%$ of the deep eigenvectors (the overlaps are in Appendix~\ref{app:ho}), and the few degrees are the on-shell consequence of this deficit. The same deficit shows below threshold: with the Coulomb force on, the projection of the oscillator functions leaves the $\alpha\alpha$ pair with a bound state at $-0.31$~MeV, where the projection of the dressed eigenstates leaves none and $^{8}$Be is the resonance of Table~\ref{tab:coulon}; with the Coulomb force off it moves the shallow state of the projected pair from $-2.25$ to $-2.51$~MeV. In $L=2$ the error is larger, above seven degrees at $5$~MeV, above the $^{8}$Be $2^{+}$ resonance, and it again vanishes by $20$~MeV (Table~\ref{tab:conv2b}).

Table~\ref{tab:phaseeq} gives the phases themselves, computed with the Fock block, without the projection and with the dressed eigenstate projected: the two coincide, in $\alpha p$ and in $\alpha\alpha$ $L=0$ and $L=2$ at every energy of the table, as the orthogonality of the projection requires.  Its difference columns are the non-eigenstate cases: a two-pole vector for the $\alpha\alpha$ $L=0$ forbidden states whose eigenvalues lie $40$ and $13$~keV from the dressed eigenstates; the dressed $\alpha p$ state projected from the nuclear-only problem, without the Coulomb interaction; and the undressed (Coulomb-subtracted) $\alpha\alpha$ eigenstates projected with the Coulomb interaction included; the oscillator case is in Table~\ref{tab:conv2b}.  The undressed shift is of the order of a degree in $L=0$, largest at $1$--$2$~MeV and below the oscillator shift up to $10$~MeV, and it is small in $L=2$, where the centrifugal barrier suppresses the Coulomb distortion of the deep orbit. The on-shell effect of projecting a vector that is not an eigenstate is determined not by its overlap deficit alone but by the extent to which its deviation affects the scattering wave function at that energy. The projection of a true eigenvector is therefore phase-equivalent to the unprojected separable potential at positive energy, within the accuracy of the fit; the two phases differ only by the Levinson multiple of $\pi$ at threshold. In this respect the strict projection is analogous to the supersymmetric removal of a bound state, which also preserves the phases~\cite{Deltuva2026}: the projected, the unprojected and the supersymmetrically transformed interactions have the same two-body phases~\cite{Baye1987} and differ only off shell, at the negative energies which the three-body kernel samples, and the three-body energies are what distinguishes them. A state which is an eigenvector only to within $40$~keV, like the $40$-keV column of Table~\ref{tab:phaseeq}, leaves the phase unchanged; the same dressed $\alpha p$ state projected from the nuclear-only problem, of which it is not an eigenvector, shifts it only slightly. The effect increases with the deviation from an eigenvector, and the oscillator columns of Table~\ref{tab:conv2b} give its magnitude for the oscillator functions.

\begin{table}[htbp]
\caption{The on-shell effect of projecting a vector which is not an eigenstate, as differences on $-$ off of the Coulomb-modified nuclear phase $\delta_{SC}$ (degrees) in the last four columns; the first columns give the phase itself without (off) and with (on) the strict Feshbach--Schur projection of the dressed forbidden eigenstate, identical by the eigenstate condition. Differences of a thousandth of a degree or less are zero within the numerical precision of the phase.}
\label{tab:phaseeq}
\begin{center}
\footnotesize
\resizebox{\textwidth}{!}{%
\begin{tabular}{c|cc|cc|cc|cccc}
\toprule
$E_{\rm cm}$ & \multicolumn{2}{c|}{$\alpha p$ $^{2}S_{1/2}$} & \multicolumn{2}{c|}{$\alpha\alpha$ $L=0$} & \multicolumn{2}{c|}{$\alpha\alpha$ $L=2$} & $L=0$, $40$~keV & $\alpha p$, nucl. & $L=0$, undr. & $L=2$, undr. \\
(MeV) & off & on & off & on & off & on & on $-$ off & on $-$ off & on $-$ off & on $-$ off \\
\midrule
 $0.1$ & $-0.3947$ & $-0.3947$ & $0.0579$   & $0.0579$   & $0.0000$   & $0.0000$   & $+0.0001$ & $+0.050$ & $-0.0629$ & $+0.0000$ \\
 $0.5$ & $-8.0468$ & $-8.0468$ & $-10.0037$ & $-10.0037$ & $0.0187$   & $0.0187$   & $+0.0001$ & $+0.103$ & $+1.2194$ & $+0.0001$ \\
 $1$   & $-16.8334$ & $-16.8334$ & $-32.9724$ & $-32.9724$ & $0.4778$  & $0.4778$  & $+0.0002$ & $+0.132$ & $+1.6038$ & $+0.0044$ \\
 $2$   & $-29.7208$ & $-29.7208$ & $-68.6384$ & $-68.6384$ & $9.7099$  & $9.7099$  & $+0.0002$ & $+0.157$ & $+1.5130$ & $+0.1551$ \\
 $5$   & $-53.3540$ & $-53.3540$ & $51.9577$  & $51.9577$  & $-63.5129$ & $-63.5129$ & $+0.0001$ & $+0.159$ & $+1.2116$ & $+0.3940$ \\
 $10$  & $-75.5326$ & $-75.5326$ & $0.9191$   & $0.9191$   & $-79.3728$ & $-79.3728$ & $0.0000$  & $+0.130$ & $+0.9964$ & $+0.1600$ \\
 $20$  & $81.3524$  & $81.3524$  & $-52.7717$ & $-52.7717$ & $70.8077$  & $70.8077$  & $0.0000$  & $+0.086$ & $+0.7449$ & $+0.0865$ \\
\bottomrule
\end{tabular}}
\end{center}
\end{table}

\section{Extension to positive energy}
\label{sec:scattering}

The representation~(\ref{eq:fock}) is an expansion of the Coulomb interaction at negative energies, and what a finite truncation does at positive energies was given above. Two parts of the extension to positive energy are two-body statements and are given here. For two-body scattering the Coulomb-modified phase is then obtained with the singularity treated exactly, and the construction is tested against known phases. For the strip of positive pair energies which opens in a three-body kernel above the breakup threshold, a two-body test establishes what a finite Fock truncation is there and how its phase is renormalized. Elastic three-body scattering below breakup, where every pair energy the kernel samples is negative, is the application that follows.

\subsection{Two-body scattering with the singularity treated exactly}
\label{sec:twobodyscat}

Let the pair interaction be $V = V_{C} + V_{s}$, with the point Coulomb force $V_{C}$ and the separable nuclear part $V_{s}$ of Eq.~(\ref{eq:est}). The two-potential decomposition, Eq.~(\ref{eq:tscreen}) with the point Coulomb force in the place of the screened one, gives the exact $t$~matrix
\begin{equation}
t(E) \;=\; t_{C}(E)
\;+\; \bigl(1+t_{C}G_{0}\bigr)\,
\sum_{ij}\ket{\chi_{i}}\,
\bigl[\bm\Lambda^{-1}-\bm{\mathcal D}_{C}(E)\bigr]^{-1}_{ij}
\bra{\chi_{j}}\,\bigl(1+G_{0}t_{C}\bigr),
\label{eq:twopot}
\end{equation}
where $t_{C}$ is the pure Coulomb $t$~matrix.  The only change with respect to the free-space separable formula is the replacement of the free resolvent by the Coulomb one in the propagator matrix,
\begin{equation}
\bigl[\bm{\mathcal D}_{C}(E)\bigr]_{ij}
\;=\; \bra{\chi_{i}}\,G_{C}(E+i\epsilon)\,\ket{\chi_{j}},
\qquad G_{C}(z) = (z - H_{0} - V_{C})^{-1}.
\label{eq:DC}
\end{equation}
On shell the physical amplitude separates into the Coulomb amplitude plus a Coulomb-distorted separable remainder, and the phase shift has the standard form $\delta_{L} = \sigma_{L} + \delta_{SC,L}$: the pure Coulomb phase $\sigma_{L}$ plus the Coulomb-modified nuclear phase
\begin{equation}
\tan\delta_{SC,L}
\;=\; -\frac{2\mu}{k}\;
\bm v^{T}\bigl[\bm\Lambda^{-1}-\bm{\mathcal D}^{C}(E_{k})\bigr]^{-1}\bm v,
\qquad
v_{i}=\braket{F_{L}(\eta,k\,\cdot)|\tilde\chi_{i}},
\label{eq:deltaSC}
\end{equation}
in the standing-wave convention, with $\bm{\mathcal D}^{C}$ the principal-value Coulomb propagator matrix, the real part of $\bm{\mathcal D}_{C}$ of Eq.~(\ref{eq:DC}) (the superscript marks the principal-value matrix throughout), and
\begin{equation}
\tilde\chi_{i}(r)=\sqrt{2/\pi}\int_{0}^{\infty}dp\,\hat\jmath_{L}(pr)\,p\,\chi_{i}(p)
\label{eq:radial}
\end{equation}
the radial transform of the form factor, Eq.~(\ref{eq:deltaSC}) giving $\delta_{SC,L}$ modulo $180^{\circ}$, quoted in the tables in $(-90^{\circ},90^{\circ}]$; $F_{L}$ and $G_{L}$ are the regular and irregular Coulomb functions and $\hat\jmath_{L}$, $\hat n_{L}$ the Riccati--Bessel functions, $\hat n_{L}$ in the convention $\hat n_{L}(kr)=G_{L}(0,kr)$.  Nothing is fitted; for $\eta\to0$ Eq.~(\ref{eq:deltaSC}) reduces to the free-space relation used to verify the separable potentials themselves. With the screened force $V^{R}_{C}$ in the place of $V_{C}$ the same formula holds with the regular and irregular solutions of the screened Coulomb problem in the place of $F_{L}$ and $G_{L}$, and gives the difference $\delta^{R}_{L}-\delta^{R}_{C,L}$ of Eq.~(\ref{eq:screenren}) directly; the construction of this section is the limit $R\to\infty$ of the screening method taken analytically, the Coulomb functions containing the logarithm which the screened solutions accumulate as $R$ increases.

The on-shell Coulomb logarithm, which no finite Fock expansion and no separable form factor can represent, is contained in the partial-wave Coulomb Green's function,
\begin{equation}
G_{C,L}(r,r';k) \;=\;
-\frac{2\mu}{k}\;F_{L}(\eta,kr_{<})\,
\bigl[G_{L}(\eta,kr_{>})+iF_{L}(\eta,kr_{>})\bigr],
\label{eq:GCcoord}
\end{equation}
known in closed form~\cite{Schwinger1964,Hostler1964,vanHaeringenBook}. The free part of the propagator matrix is evaluated in momentum space without approximation, and coordinate space enters only through the difference of the Coulomb and free $F_{L}G_{L}$ kernels, the standing-wave part of Eq.~(\ref{eq:GCcoord}) minus its free counterpart,
\begin{multline}
\bigl[\bm{\mathcal D}^{C}(E_{k})-\bm{\mathcal D}(E_{k})\bigr]_{ij}
=-\frac{2\mu}{k}\int_{0}^{\infty}\!dr\int_{0}^{\infty}\!dr'\,
\tilde\chi_{i}(r)\\
\times\bigl[F_{L}(\eta,kr_{<})\,G_{L}(\eta,kr_{>})
-\hat\jmath_{L}(kr_{<})\,\hat n_{L}(kr_{>})\bigr]\,\tilde\chi_{j}(r'),
\label{eq:diffkernel}
\end{multline}
a double radial quadrature with the closed-form radial transforms~(\ref{eq:radial}) of the form factors; the overlaps $v_{i}$ are single quadratures with $F_{L}(\eta,kr)$. The bracket is finite at $r=r'$ and vanishes identically as $\eta\to0$, so the quadrature errors cancel in the free limit, which is exact by construction; at finite $\eta$ it is confirmed by the convergence of the renormalized Fock phase to it (Sec.~\ref{sec:abovebreak}). No screening radius appears, and the logarithmic $N_{C}$ dependence of the truncated representation does not appear. In momentum space the same statement is that the Coulomb $t$~matrix and Green's function have forms in which the on-shell singularity is separated from a regular remainder~\cite{StorozhenkoShadchin}; the separable nuclear force couples only to the remainder, of which the Fock modes are an expansion. This separation is used for the positive pair energies above the breakup threshold.

The two-potential formula with the exact partial-wave Coulomb Green's function is well known~\cite{vanHaeringen1982,vanHaeringenBook}; the only new element here is its application to the separable potentials of this paper (Ref.~\cite{NishonovValidation}) and with the strict projection over the forbidden block, needed for the two-body comparisons. In that combination the projection is a Schur complement over the forbidden block of the matrix in Eq.~(\ref{eq:deltaSC}), with $\mathsf{A}$ and $\mathsf{F}$ labelling the allowed and the forbidden blocks, and must act also on the overlap vector,
\begin{equation}
\tan\delta_{SC,L}
=-\frac{2\mu}{k}\Bigl[\bm v_{\rm eff}^{T}\,\bm{\mathcal S}^{-1}\,\bm v_{\rm eff}
-\bm v_{\mathsf{F}}^{T}\bigl(\bm{\mathcal D}^{C}_{\mathsf{FF}}\bigr)^{-1}\bm v_{\mathsf{F}}\Bigr],
\qquad
\bm v_{\rm eff}=\bm v_{\mathsf{A}}-\bm{\mathcal D}^{C}_{\mathsf{AF}}\bigl(\bm{\mathcal D}^{C}_{\mathsf{FF}}\bigr)^{-1}\bm v_{\mathsf{F}},
\label{eq:veff}
\end{equation}
where
\begin{equation*}
\bm{\mathcal S}=\bm\Lambda^{-1}-\bm{\mathcal D}^{C}_{\mathsf{AA}}
+\bm{\mathcal D}^{C}_{\mathsf{AF}}\bigl(\bm{\mathcal D}^{C}_{\mathsf{FF}}\bigr)^{-1}\bm{\mathcal D}^{C}_{\mathsf{FA}}
\end{equation*}
is the Schur complement of the allowed block and $\bm v_{\mathsf{A}}$, $\bm v_{\mathsf{F}}$ are the allowed and forbidden parts of the overlap vector; the second term is the $\lambda\to\infty$ limit of the forbidden block's own contribution, and at $\eta=0$ the expression reduces to the $\lambda\to\infty$ solution of the pseudopotential problem.

Three tests of Eq.~(\ref{eq:deltaSC}) are collected in Tables~\ref{tab:val2b} and~\ref{tab:coulon}. At $\eta=0$ the $pp$ $^{1}S_{0}$ separable potential reproduces the scattering length and effective range of the CD~Bonn potential it represents. With the point Coulomb interaction included, the $pp$ $^{1}S_{0}$ phases agree closely with the Nijmegen partial-wave analysis PWA93~\cite{Stoks1993} up to $100$~MeV, and the Coulomb-modified effective-range parameters, from a two-parameter fit over $E_{\rm cm}=0.01$--$0.12$~MeV, agree with the CD~Bonn values (Table~\ref{tab:coulon}), the uncertainties being the spread over subsets of the fit points. For the $\alpha\alpha$ $L=0$ separable potential, whose original potential is fitted together with its folded Coulomb force, the same construction gives the phases of Table~\ref{tab:val2b} and places the $^{8}$Be ground-state resonance $15\%$ above experiment (Table~\ref{tab:coulon}). This is the finite-rank error of the separable potential at a point of extreme threshold sensitivity, not an error of the Coulomb construction. The resonance lies less than $0.1$~MeV above threshold, so the $14$~keV excess of the fitted pole over experiment (Table~\ref{tab:coulon}; $19$~keV against the original potential), negligible at MeV energies, is the $15\%$ quoted. At $\eta=0$ the same separable potential reproduces its original potential accurately over $E_{\rm cm}=1$--$12$~MeV. The Coulomb-modified construction does not increase that residual. The local folded-Coulomb original potential itself lies close to experiment (Table~\ref{tab:coulon}).  In the $L=2$ channel the $^{8}$Be $2^{+}$ resonance lies $6\%$ above experiment, with a width larger than the experimental one.

\begin{table}[htbp]
\caption{Two-body validations of the Coulomb-modified phase, Eq.~(\ref{eq:deltaSC}): $pp$ against the CD~Bonn potential and PWA93~\cite{Stoks1993}, and the $\alpha\alpha$ phases of the separable potential with the point Coulomb force; the Coulomb-modified threshold parameters and the $^{8}$Be resonances are compared with experiment in Table~\ref{tab:coulon}. Phases in degrees, $T_{\rm lab}$ and $E_{\rm cm}$ in MeV, lengths in fm.}
\label{tab:val2b}
\begin{center}
\begin{tabular}{lccl}
\toprule
\multicolumn{4}{l}{$pp$ $^{1}S_{0}$ phase} \\
$T_{\rm lab}$ & this work & PWA93 & \\
\midrule
$5$   & $54.87$ & $54.83$ & \\
$10$  & $55.22$ & $55.22$ & \\
$25$  & $48.62$ & $48.67$ & \\
$50$  & $38.80$ & $38.93$ & \\
$100$ & $24.82$ & $24.99$ & \\
\midrule
\multicolumn{4}{l}{$\alpha\alpha$ phase} \\
$E_{\rm cm}$ & $L=0$ & $L=2$ & \\
\midrule
$0.5$ & $-10.00$ & $0.02$   & \\
$2$   & $-68.64$ & $9.71$   & \\
$5$   & $51.96$  & $-63.51$ & \\
$20$  & $-52.77$ & $70.81$  & \\
\midrule
quantity & this work & reference & \\
\midrule
$pp$ $^{1}S_{0}$, $\eta=0$: $a$, $r$      & $-17.459(1)$, $2.840(3)$ & $-17.460$, $2.844$ & original \\
\bottomrule
\end{tabular}
\end{center}
\end{table}

\subsection{Positive pair energies above the breakup threshold}
\label{sec:abovebreak}

\begin{figure}[htbp]
\begin{center}
\begin{tikzpicture}
\begin{groupplot}[group style={group size=3 by 1, horizontal sep=1.25cm},
  width=0.335\textwidth, height=5.4cm,
  tick label style={font=\scriptsize}, label style={font=\scriptsize},
  title style={font=\scriptsize}, every mark/.append style={scale=0.9},
  legend style={font=\scriptsize, draw=none, fill=none, inner sep=1pt}, legend cell align=left]
\nextgroupplot[title={$pp$ $^{1}S_{0}$}, xlabel={$T_{\rm lab}$ (MeV)}, ylabel={$\delta_{SC}$ (deg)},
  xmin=0, xmax=105, legend pos=north east]
\addplot[black] coordinates {(5,54.870) (10,55.220) (25,48.620) (50,38.800) (100,24.820)}; \addlegendentry{exact}
\addplot[black, only marks, mark=o] coordinates {(5,54.830) (10,55.220) (25,48.670) (50,38.930) (100,24.990)}; \addlegendentry{PWA93}
\addplot[black, only marks, mark=*] coordinates {(5,54.887) (25,48.620) (50,38.800)}; \addlegendentry{Fock}
\nextgroupplot[title={$\alpha p$ $^{2}S_{1/2}$}, xlabel={$E_{\rm cm}$ (MeV)}, xmode=log, log basis x=10,
  xtick={0.1,0.5,2,5,20}, xticklabels={0.1,0.5,2,5,20}, legend pos=south west]
\addplot[black] coordinates {(0.1,-0.395) (0.5,-8.047) (1,-16.833) (2,-29.721) (5,-53.354) (10,-75.533) (20,-98.648)}; \addlegendentry{exact}
\addplot[black, only marks, mark=*] coordinates {(0.5,-8.016) (2,-29.728) (5,-53.379) (20,-98.640)}; \addlegendentry{Fock}
\nextgroupplot[title={$\alpha\alpha$ $L=0$}, xlabel={$E_{\rm cm}$ (MeV)}, xmode=log, log basis x=10,
  xtick={0.1,0.5,2,5,20}, xticklabels={0.1,0.5,2,5,20}, legend pos=south west, legend style={row sep=-3pt}]
\addplot[black] coordinates {(0.107,179.404) (0.108,179.671) (0.11,179.807) (0.115,179.875) (0.12,179.887) (0.13,179.877) (0.15,179.807) (0.175,179.650) (0.2,179.407) (0.25,178.633) (0.3,177.469) (0.35,175.956) (0.4,174.161) (0.45,172.154) (0.5,169.996) (1,147.028) (2,111.362) (5,51.958) (10,0.919) (20,-52.772)}; \addlegendentry{exact}
\addplot[black, only marks, mark=*] coordinates {(0.5,169.995) (2,111.234) (5,52.001) (20,-52.747)}; \addlegendentry{Fock}
\end{groupplot}
\end{tikzpicture}
\end{center}
\caption{The Coulomb-modified nuclear phase $\delta_{SC}$ against energy: the exact construction, Eq.~(\ref{eq:deltaSC}) (lines), the renormalized Fock phase, Eq.~(\ref{eq:renphase}), at $N_{C}=320$ (filled circles), and for $pp$ the analysis PWA93~\cite{Stoks1993} (open circles). The $\alpha\alpha$ line starts above the $^{8}$Be ground-state resonance at $0.106$~MeV, across which the phase has risen through $180^{\circ}$.}
\label{fig:phases}
\end{figure}
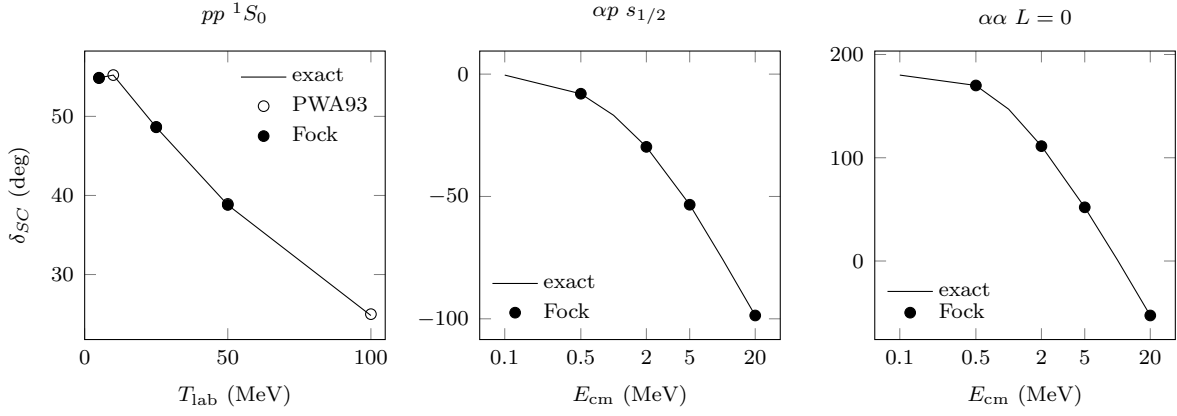

Above the breakup threshold a finite strip of positive pair energies,
\begin{equation}
q<q_{\mathrm{crit}}=\sqrt{2\bar\mu z},
\label{eq:strip}
\end{equation}
with $q$ the spectator momentum, $\bar\mu$ the spectator reduced mass and $z$ the three-body energy above the breakup threshold, opens in the kernel, and only there positive-energy Coulomb quantities appear. A two-body test determines what the Fock block does there. In the $\alpha p$ $^{2}S_{1/2}$ and $\alpha\alpha$ $L=0$ channels, at $E=0.5$--$20$~MeV and with the strict projection, the Fock phase relative to free waves deviates from the exact construction logarithmically in $N_{C}$, with the slope of Sec.~\ref{sec:boundary}: the truncation is a screened potential and the $N_{C}$ dependence its screening phase. The consistent construction is that required in any screened calculation, the total phase of nuclear plus truncated Coulomb minus the pure phase of the truncated Coulomb alone,
\begin{equation}
\delta_{SC}^{(N_{C})}=\delta_{\rm tot}^{(N_{C})}-\delta_{C}^{(N_{C})},
\label{eq:renphase}
\end{equation}
with free waves in the external states everywhere: both phases are the free-space phases of finite-rank separable potentials, Eq.~(\ref{eq:freephase}), $\delta_{\rm tot}^{(N_{C})}$ over the extended basis~(\ref{eq:basisorder}) with the strengths~(\ref{eq:Lamext}) and the strict projection where the channel has forbidden states, $\delta_{C}^{(N_{C})}$ over the $N_{C}$ Fock modes alone.  This renormalized phase converges to the exact one in both channels as $1/N_{C}$, and at $N_{C}=640$ the two coincide on the scale of the figure at every energy. Figure~\ref{fig:phases} shows the renormalized phase at $N_{C}=320$ with the exact one for the three pairs, and for $pp$ the PWA93 phases~\cite{Stoks1993}. The phases are drawn continuous modulo $180^{\circ}$ on the branch of the projected problem, in which the forbidden states are removed and Levinson's theorem gives $\delta_{SC}=0$ at threshold. On that branch the $\alpha p$ phase falls from zero through $-90^{\circ}$, and the $\alpha\alpha$ phase rises to $180^{\circ}$ across the narrow $^{8}$Be ground-state resonance and falls from there; without the projection the same phases start at the Levinson multiple, $\pi$ for $\alpha p$ and $2\pi$ for $\alpha\alpha$. The points lie on the curves also where the error of a finite truncation would be largest, through the $^{8}$Be region of the $\alpha\alpha$ phase and across the $-90^{\circ}$ crossing of the $\alpha p$ phase. The renormalized phase is free of the logarithm.

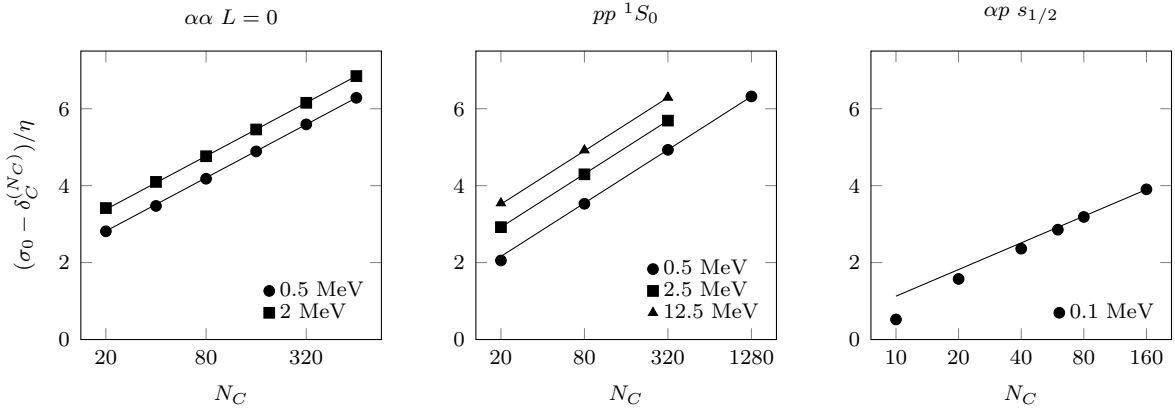
\begin{figure}[tb]
	\begin{center}
		\begin{tikzpicture}
			\begin{groupplot}[group style={group size=3 by 1, horizontal sep=1.25cm},
				width=0.335\textwidth, height=5.4cm, xmode=log, log basis x=10, xlabel={$N_{C}$},
				ymin=0, ymax=7.5,
				tick label style={font=\scriptsize}, label style={font=\scriptsize},
				title style={font=\scriptsize}, every mark/.append style={scale=0.9},
				legend style={font=\scriptsize, draw=none, fill=none, inner sep=1pt, row sep=-3pt}, legend cell align=left]
				\nextgroupplot[title={$\alpha\alpha$ $L=0$}, xtick={20,80,320}, xticklabels={20,80,320}, ylabel={$(\sigma_{0}-\delta_{C}^{(N_{C})})/\eta$}, legend pos=south east]
				\addplot[black, forget plot] coordinates {(20,2.819) (40,3.512) (80,4.205) (160,4.899) (320,5.592) (640,6.285)};
				\addplot[black, only marks, mark=*] coordinates {(20,2.813) (40,3.473) (80,4.180) (160,4.891) (320,5.593) (640,6.284)}; \addlegendentry{$0.5$~MeV}
				\addplot[black, forget plot] coordinates {(20,3.383) (40,4.076) (80,4.769) (160,5.462) (320,6.156) (640,6.849)};
				\addplot[black, only marks, mark=square*] coordinates {(20,3.417) (40,4.098) (80,4.765) (160,5.461) (320,6.155) (640,6.848)}; \addlegendentry{$2$~MeV}
				\nextgroupplot[title={$pp$ $^{1}S_{0}$}, xtick={20,80,320,1280}, xticklabels={20,80,320,1280}, legend pos=south east]
				\addplot[black, forget plot] coordinates {(20,2.161) (80,3.547) (320,4.933) (1280,6.320)};
				\addplot[black, only marks, mark=*] coordinates {(20,2.055) (80,3.529) (320,4.931) (1280,6.320)}; \addlegendentry{$0.5$~MeV}
				\addplot[black, forget plot] coordinates {(20,2.919) (80,4.305) (320,5.691)};
				\addplot[black, only marks, mark=square*] coordinates {(20,2.921) (80,4.295) (320,5.691)}; \addlegendentry{$2.5$~MeV}
				\addplot[black, forget plot] coordinates {(20,3.519) (80,4.905) (320,6.291)};
				\addplot[black, only marks, mark=triangle*] coordinates {(20,3.546) (80,4.922) (320,6.291)}; \addlegendentry{$12.5$~MeV}
				\nextgroupplot[title={$\alpha p$ $^{2}S_{1/2}$}, xtick={10,20,40,80,160}, xticklabels={10,20,40,80,160}, legend pos=south east]
				\addplot[black, forget plot] coordinates {(10,1.127) (20,1.820) (40,2.513) (60,2.919) (80,3.206) (160,3.900)};
				\addplot[black, only marks, mark=*] coordinates {(10,0.522) (20,1.575) (40,2.362) (60,2.856) (80,3.187) (160,3.905)}; \addlegendentry{$0.1$~MeV}
			\end{groupplot}
		\end{tikzpicture}
	\end{center}
	\caption{The screening logarithm of the truncated Fock block, $(\sigma_{0}-\delta_{C}^{(N_{C})})/\eta$ from the computed pure Coulomb phase (points), as a function of the rank $N_{C}$, and the law $\ln(2kR_{N})$ of Eqs.~(\ref{eq:screenphase}) and~(\ref{eq:RN}) (lines).}
	\label{fig:screenlaw}
\end{figure}

The pure Coulomb phase of the rank-$N_{C}$ Fock potential obeys
\begin{equation}
\delta^{(N_{C})}_{C}=\sigma_{0}(\eta)
-\eta\ln\!\Bigl(\frac{4kN_{C}\beta_{C}}{\beta_{C}^{2}+k^{2}}\Bigr)
+O(1/N_{C}),
\label{eq:screenphase}
\end{equation}
so the screening radius of the truncation is
\begin{equation}
R_{N}=\frac{2N_{C}\beta_{C}}{\beta_{C}^{2}+k^{2}},
\qquad 2kR_{N}=2N_{C}\sin\theta_{k},
\qquad k=\beta_{C}\tan(\theta_{k}/2),
\label{eq:RN}
\end{equation}
with $\theta_{k}$ the Fock angle, the polar angle of the stereographic map~(\ref{eq:stereo}) on the three-sphere, given by Eq.~(\ref{eq:chi}) of Appendix~\ref{app:derivation} at $p=k$, so that $\tan(\theta_{k}/2)=k/\beta_{C}$. Equation~(\ref{eq:screenphase}) is derived at leading order in $1/N_{C}$ at the end of Appendix~\ref{app:derivation}, from the partial sum of the Fock series, and the computed phases confirm it (Fig.~\ref{fig:screenlaw}). Its form has a simple origin: the truncation resolves the Fock angle to about $1/N_{C}$, which at the momentum $k$ is the resolution $(\beta_{C}^{2}+k^{2})/2\beta_{C}N_{C}$, whose inverse is $R_{N}$. 
The law holds very well for $N_{C}\ge320$ in both channels and for $\beta_{C}=0.8$--$1.5$~fm$^{-1}$, with the same screening logarithm $\ln(2kR_{N})=\ln(2N_{C}\sin\theta_{k})$ of Eq.~(\ref{eq:RN}), a function of $k/\beta_{C}$ only; at $N_{C}=160$ the law holds closely at all but the lowest $\alpha\alpha$ energy $2$~MeV. Figure~\ref{fig:screenlaw} shows the law against the computed phase of the truncated block: for $\alpha\alpha$ at $0.5$ and $2$~MeV and $pp$ at $0.5$--$12.5$~MeV the points approach the lines as $1/N_{C}$ and coincide with them for $N_{C}\ge320$, while the $\alpha p$ phase at $0.1$~MeV, the non-convergent row of Table~\ref{tab:boundary}, approaches its line from below, deviating by thirty degrees at $N_{C}=10$ and reaching its line at $N_{C}=160$. The figure shows the phase as the screening logarithm $(\sigma_{0}-\delta_{C})/\eta$, in which every line has unit slope in $\ln N_{C}$ and the intercept $\ln[4k\beta_{C}/(\beta_{C}^{2}+k^{2})]$ contains the energy dependence. On the strip, therefore, the Fock representation acts as the screening-and-renormalization scheme of Ref.~\cite{Alt1978} in separable and analytic form: the screening is the truncation of the rank, with the radius $R_{N}$ which varies with the pair momentum, and the renormalization is the subtraction of the pure Coulomb phase of the same truncation, convergent in $N_{C}$ as the screened limit is in $R$. The screening radius differs for every pair energy of the strip; the use of this result in the three-body scattering equations, together with the renormalization question which Refs.~\cite{Deltuva2024,Witala2024} debate for the screened case, is left to the three-body scattering work that follows.

\section{Summary and outlook}
\label{sec:summary}
\label{sec:eft}
The unscreened Coulomb interaction is exactly separable by the $O(4)$ symmetry of Fock, with a truncation that converges at negative energies, and is added to the separable nuclear potential of a charged pair as a block of separable terms with analytic form factors and strengths; the Feshbach--Schur projection then acts on the Coulomb-dressed forbidden state which the eigenstate condition of Ref.~\cite{Nishonov2026a} requires, and which the extended separable problem provides.

The block was tested against known quantities, the $pp$ scattering length and $^{1}S_{0}$ phases and the $\alpha p$ and $\alpha\alpha$ Coulomb-modified phases, none of which is a result of this paper. Compared with the other separable Coulomb treatments it reproduces the exact off-shell $t$~matrix at the percent level, where the Coulomb-distorted form factors deviate by $12$--$22\%$, and the displacement of the $\alpha p$ forbidden state, where the distortion method overestimates it by a factor of two, the double counting of its propagator matrix, Eq.~(\ref{eq:distdiff}). The two-body Coulomb-modified phase is computed with the singularity carried exactly in the Coulomb Green's function; the $^{8}$Be ground-state resonance comes out above experiment, a property of the $\alpha\alpha$ separable potential, and the $2^{+}$ above it as in the BFW potential itself. At positive energy a finite truncation is a screened Coulomb potential with the radius of Eq.~(\ref{eq:RN}) and the screening phase of Eq.~(\ref{eq:screenphase}); its renormalized phase converges to the exact one as $1/N_{C}$, so that for the pair amplitudes above the three-body breakup threshold the representation would act as the screening-and-renormalization scheme in analytic form, with a radius which varies with the pair momentum.

For the eigenstate condition, the projection acts on the forbidden bound state alone only for the eigenstate of the Hamiltonian in use, the Coulomb-dressed state. On shell the oscillator function changes the $\alpha\alpha$ phases by a few degrees at the lowest energies, the undressed eigenstate by a fraction of that; the three-body consequences, $1.63$~MeV in $^{12}$C for the undressed eigenstate and $2.29$~MeV for the oscillator function, are given in Ref.~\cite{FockII}.

The representation does not expand the on-shell Coulomb singularity, nor will any finite truncation: the $N_{C}$ dependence of Sec.~\ref{sec:boundary} is the screening phase of the truncation, which is removed by the subtraction, not by additional terms. A separable potential with the Coulomb force inside its fit remains incorrect on shell.  With the Coulomb interaction included, the required forbidden eigenvector is the Coulomb-dressed one, which the Fock block provides.

The interactions of pionless and halo effective field theory are contact terms, separable potentials of finite rank, and their three-body equations have the form of those of Ref.~\cite{FockII}, with form factors fixed by the power counting~\cite{HammerJiPhillips2017}. The Coulomb force is treated there exactly in the two-body sector, for $pp$~\cite{KongRavndal2000} and for $\alpha\alpha$~\cite{Higa2008}, and by screening or direct integration in the three-body sector, in $pd$ scattering and $^{3}$He~\cite{KonigHammer2011,Vanasse2014}. The Fock block can be added to these kernels without changing their separable structure. At negative pair energies it is exact, with analytic strengths and no screening. At positive pair energies it is the screened potential of Sec.~\ref{sec:boundary}, with a known radius and screening phase.

\appendix

\section{Derivation of the Fock representation}
\label{app:derivation}

The separable form of the representation is derived in four steps: the stereographic map of the momentum space onto the sphere, the expansion of the sphere kernel in $S^{3}$ harmonics, its partial-wave reduction, and the normalization of the strengths.

\subsection*{The stereographic map and the chord relation}

Fix a scale $\beta_{C}>0$ and map each momentum $\bm p\in\mathbb R^{3}$ to a unit vector on the three-sphere,
\begin{equation}
u(\bm p) \;=\;
\Bigl(\frac{2\beta_{C}\,\bm p}{\beta_{C}^{2}+p^{2}},\;
\frac{\beta_{C}^{2}-p^{2}}{\beta_{C}^{2}+p^{2}}\Bigr),
\qquad |u|=1 .
\label{eq:stereo}
\end{equation}
Direct computation expresses the momentum transfer through the chord between the images of $\bm p$ and $\bm p'$ on the sphere,
\begin{equation}
|u(\bm p)-u(\bm p')|^{2}
\;=\;
\frac{4\beta_{C}^{2}\,|\bm p-\bm p'|^{2}}
{(\beta_{C}^{2}+p^{2})(\beta_{C}^{2}+p'^{2})} .
\label{eq:chord}
\end{equation}
In the original treatment of Fock $\beta_{C}$ is tied to the energy, $\beta_{C}=\sqrt{-2\mu E}$, which transforms the Schr\"odinger equation of the hydrogen atom into an integral equation on $S^{3}$ with manifest $O(4)$ symmetry. For the present purpose the relation~(\ref{eq:chord}) is purely algebraic and holds for any fixed $\beta_{C}$; this is why the expansion of the potential below does not depend on the energy and why $\beta_{C}$ is an auxiliary scale.

\subsection*{Expansion of the sphere kernel}

The momentum-space Coulomb kernel is, in the units $\hbar=c=1$,
\begin{equation}
\braket{\bm p|V_{C}|\bm p'}=\frac{Z_{1}Z_{2}\,e^{2}}{2\pi^{2}\,|\bm p-\bm p'|^{2}}.
\label{eq:VCp}
\end{equation}
With the chord relation~(\ref{eq:chord}), Eq.~(\ref{eq:VCp}) becomes
\begin{equation}
\braket{\bm p|V_{C}|\bm p'}
=\frac{Z_{1}Z_{2}\,e^{2}}{2\pi^{2}}\,
\frac{4\beta_{C}^{2}}{(\beta_{C}^{2}+p^{2})(\beta_{C}^{2}+p'^{2})}\;
\frac{1}{|u-u'|^{2}} .
\label{eq:VonSphere}
\end{equation}
The sphere kernel is zonal, $|u-u'|^{2}=2(1-\cos\omega)$ with $\omega$ the $S^{3}$ angle between $u$ and $u'$, and its expansion follows from the generating function of the Gegenbauer polynomials,
\begin{equation}
\sum_{n\ge0}C^{(1)}_{n}(x)\,t^{n}=\frac{1}{1-2xt+t^{2}},
\label{eq:genfun}
\end{equation}
taken in the limit $t\to1^{-}$, $t$ approaching $1$ from below within the disc $|t|<1$ in which the series converges; at $t=1$ itself the terms $C^{(1)}_{n}(\cos\omega)=\sin[(n+1)\omega]/\sin\omega$ do not decrease and the series is summable only in Abel's sense, its Abel sum being the left-hand side:
\begin{equation}
\frac{1}{|u-u'|^{2}}
=\frac{1}{2(1-\cos\omega)}
=\sum_{n=0}^{\infty} C^{(1)}_{n}(\cos\omega) .
\label{eq:zonal}
\end{equation}
The expansion is used inside the partial-wave kernel, where after the projection on $L$ it becomes the convergent series of Eq.~(\ref{eq:chebylog}) for the logarithmic kernel, and its truncation at rank $N_{C}$ is analysed at the end of this appendix.
The addition theorem for the $S^{3}$ harmonics,
\begin{equation}
C^{(1)}_{N}(u\!\cdot\!u')
=\frac{2\pi^{2}}{N+1}\sum_{L=0}^{N}\sum_{m=-L}^{L}
Y_{NLm}(u)\,Y^{*}_{NLm}(u'),
\label{eq:addition}
\end{equation}
with $N$ the principal number of the harmonic, $L$ its orbital momentum, and $Y_{NLm}$ normalized to unity on $S^{3}$, turns Eq.~(\ref{eq:zonal}) into a sum of separable terms with the strengths $2\pi^{2}/(N+1)$: the Coulomb kernel is diagonal in the $S^{3}$ harmonics with eigenvalues which fall as $1/(N+1)$. This is the diagonalization of Fock, applied to the potential rather than to the bound-state spectrum.

\subsection*{Partial-wave reduction}

The $S^{3}$ harmonics factorize as
\begin{equation}
Y_{NLm}(u)=A_{NL}\,\sin^{L}\!\theta\;C^{(L+1)}_{N-L}(\cos\theta)\,Y_{Lm}(\hat
p),
\label{eq:Yfactor}
\end{equation}
with $A_{NL}$ the normalization constant of Eq.~(\ref{eq:Nnorm}) below, fixed by the unit norm of $Y_{NLm}$ under the measure $\sin^{2}\!\theta\,d\theta\,d\Omega_{\hat p}$ of the sphere, and where $\theta$ is the polar $S^{3}$ angle of the map~(\ref{eq:stereo}), the Fock angle $\theta_{k}$ of Eq.~(\ref{eq:RN}) for $p=k$,
\begin{equation}
\cos\theta=\frac{\beta_{C}^{2}-p^{2}}{\beta_{C}^{2}+p^{2}}=x_{C},
\qquad
\sin\theta=\frac{2\beta_{C} p}{\beta_{C}^{2}+p^{2}},
\qquad
\tan\frac{\theta}{2}=\frac{p}{\beta_{C}},
\label{eq:chi}
\end{equation}
the last by the half-angle identity. At $p=k$ this angle is the $\theta_{k}$ of Eq.~(\ref{eq:RN}), whose two relations follow from it. After collecting the factors $(\beta_{C}^{2}+p^{2})$ from Eqs.~(\ref{eq:VonSphere}) and~(\ref{eq:chi}), the projection of the Coulomb kernel on the orbital momentum $L$ becomes
\begin{equation}
V_{C,L}(p,p') = \sum_{n=0}^{\infty}
\Phi_{nL}(p)\,\lambda_{nL}\,\Phi_{nL}(p'),
\qquad
\Phi_{nL}(p)=\frac{p^{L}\,C^{(L+1)}_{n}(x_{C})}{(\beta_{C}^{2}+p^{2})^{L+1}},
\label{eq:PWreduced}
\end{equation}
with the strengths
\begin{align}
\lambda_{nL}
&=A^{2}_{n+L,L}\,(2\beta_{C})^{2L}\,\frac{4\beta_{C}^{2}\,Z_{1}Z_{2}e^{2}}{n+L+1}
=\frac{2^{4L+3}\,Z_{1}Z_{2}\,e^{2}\,\beta_{C}^{2L+2}\,n!\,(L!)^{2}}{\pi\,(n+2L+1)!},
\label{eq:lamnL-derived}\\
A^{2}_{NL}&=\frac{2^{2L+1}\,(N+1)\,(N-L)!\,(L!)^{2}}{\pi\,(N+L+1)!},
\label{eq:Nnorm}
\end{align}
in which the Gegenbauer index $n=N-L$ counts from the lowest harmonic of the partial wave $L$; the eigenvalue $2\pi^{2}/(n+L+1)$ of the harmonic of principal number $N=n+L$ cancels the $1/(2\pi^{2})$ of the kernel~(\ref{eq:VonSphere}), the square of the constant $A_{NL}$ of Eq.~(\ref{eq:Yfactor}) at $N=n+L$, given in Eq.~(\ref{eq:Nnorm}), enters as $A^{2}_{n+L,L}$ (the Gegenbauer norm, the source of the $1/\pi$), $(2\beta_{C})^{2L}$ comes from $\sin^{L}\!\theta$ of Eq.~(\ref{eq:chi}) and $4\beta_{C}^{2}$ from the chord relation~(\ref{eq:chord}); the partial-wave amplitude $V_{C,L}$ is read off from the expansion of the kernel in $\sum_{m}Y_{Lm}(\hat p)Y^{*}_{Lm}(\hat p')$.  This is Eq.~(\ref{eq:lamnL}) of the main text.

\subsection*{Normalization of the strengths}

The overall constant is fixed by the normalization of the harmonics in Eq.~(\ref{eq:lamnL-derived}) in the measure convention used here for the propagator matrix over the extended basis~(\ref{eq:basisorder}),
\begin{equation}
\tilde{\mathcal D}_{\rho\rho'}(E)=\int_{0}^{\infty}dp\,
\frac{p^{2}\,g_{\rho}(p)\,g_{\rho'}(p)}{E-p^{2}/2\mu},
\label{eq:measure}
\end{equation}
where $g_{\rho}$ is any member of that basis, a nuclear form factor $\chi_{i}$, a Fock mode $\Phi_{n}$ or a forbidden state $\phi_{f}$, without any factors of $2\pi$. In this convention the strengths~(\ref{eq:lamnL-derived}) hold for every member of the extended basis. For $L=0$, Eq.~(\ref{eq:lamnL-derived}) reduces to Eq.~(\ref{eq:lamn}) of the main text. The normalization is confirmed by the off-shell benchmark of Sec.~\ref{sec:offshell} and by the attractive-sign block, which reproduces the hydrogen-like spectrum of the $\alpha p$ and $\alpha\alpha$ analogues at $N_{C}=80$ with $\beta_{C}$ at the Bohr momentum.

\subsection*{Origin of the screening law}

The law~(\ref{eq:screenphase}) follows at leading order in $1/N_{C}$ from the partial sum of the series~(\ref{eq:zonal}). With the same angle~(\ref{eq:chi}),
\begin{equation}
\tan\frac{\theta}{2}=\frac{p}{\beta_{C}},\qquad
\Phi_{n}(p)=\frac{\sin[(n+1)\theta]}{2\beta_{C}\,p},
\label{eq:phitheta}
\end{equation}
the strengths~(\ref{eq:lamn}) turn Eq.~(\ref{eq:fock}) into
\begin{equation}
V_{C,0}(p,p')=\frac{2Z_{1}Z_{2}e^{2}}{\pi\,p\,p'}\sum_{m=1}^{\infty}
\frac{\sin(m\theta)\sin(m\theta')}{m}
=\frac{Z_{1}Z_{2}e^{2}}{\pi\,p\,p'}\,\ln\Bigl|\frac{p+p'}{p-p'}\Bigr|,
\label{eq:chebylog}
\end{equation}
the $L=0$ Coulomb kernel, since
\begin{equation}
\sum_{m=1}^{\infty}\frac{\sin(m\theta)\sin(m\theta')}{m}
=\frac12\ln\left|\frac{\sin[(\theta+\theta')/2]}{\sin[(\theta-\theta')/2]}\right|
\label{eq:logsum}
\end{equation}
and the half-angle sines are proportional to $p\pm p'$. The rank-$N_{C}$ truncation is the partial sum of this series for the logarithmic singularity at $p=p'$. In the difference angle $\Delta\theta=\theta-\theta'$ the partial sum behaves as
\begin{equation}
\sum_{m=1}^{N_{C}}\frac{\cos(m\,\Delta\theta)}{m}
=\begin{cases}
-\ln|\Delta\theta|+O\bigl(1/(N_{C}\Delta\theta)\bigr), & \Delta\theta\gg1/N_{C},\\[3pt]
\ln N_{C}+\gamma_{\scriptscriptstyle\mathrm{E}}+O(1/N_{C}), & \Delta\theta\to0,
\end{cases}
\label{eq:partialsum}
\end{equation}
with $\gamma_{\scriptscriptstyle\mathrm{E}}$ Euler's constant, the second line being the harmonic number; the two forms coincide at
\begin{equation}
\Delta\theta_{c}=\frac{e^{-\gamma_{\scriptscriptstyle\mathrm{E}}}}{N_{C}},
\label{eq:phic}
\end{equation}
so the truncation cuts the logarithm off at that angular distance, the remainder being the oscillatory tail of the first line, of order $1/(N_{C}\Delta\theta)$.  Since $d\theta/dp=2\beta_{C}/(\beta_{C}^{2}+p^{2})$, the cutoff at the momentum
$k$ is
\begin{equation}
\Delta p_{c}=\frac{e^{-\gamma_{\scriptscriptstyle\mathrm{E}}}\,(\beta_{C}^{2}+k^{2})}{2\beta_{C}N_{C}}
=\frac{e^{-\gamma_{\scriptscriptstyle\mathrm{E}}}}{R_{N}},
\label{eq:dpc}
\end{equation}
with the $R_{N}$ of Eq.~(\ref{eq:RN}). A Coulomb potential with exponential screening at the radius $R$ has the same kernel with the logarithm cut off at $|p-p'|=1/R$, so the truncation corresponds to
\begin{equation}
R=e^{\gamma_{\scriptscriptstyle\mathrm{E}}}R_{N};
\label{eq:Rmatch}
\end{equation}
for the screening $e^{-(r/R)^{s}}$ the renormalization phase of Refs.~\cite{Alt1978,Taylor1974} is
\begin{equation}
\phi_{R}(k)=-\eta\,[\ln(2kR)-\gamma_{\scriptscriptstyle\mathrm{E}}/s],
\label{eq:aszphase}
\end{equation}
which follows from the phase accumulated by the screened tail beyond a radius $r_{0}\ll R$,
\begin{equation}
-\eta\int_{r_{0}}^{\infty}e^{-(r/R)^{s}}\,\frac{dr}{r}
=-\eta\Bigl[\ln\frac{R}{r_{0}}-\frac{\gamma_{\scriptscriptstyle\mathrm{E}}}{s}\Bigr],
\label{eq:tailphase}
\end{equation}
added to the Coulomb phase $\sigma_{0}-\eta\ln(2kr_{0})$ at $r_{0}$; for exponential screening, $s=1$, the Euler constants of Eqs.~(\ref{eq:phic}) and~(\ref{eq:aszphase}) cancel, and the phase of the truncated block is
\begin{equation}
\delta_{C}^{(N_{C})}=\sigma_{0}+\phi_{R}(k)\big|_{R=e^{\gamma_{\scriptscriptstyle\mathrm{E}}}R_{N}}
=\sigma_{0}-\eta\ln(2kR_{N})+O(1/N_{C}),
\label{eq:lawderived}
\end{equation}
Eq.~(\ref{eq:screenphase}) with no additive constant. The identification of the truncated series with exponential screening rests on their common behaviour at the diagonal; the shapes of the two cutoffs differ away from it, and the absence of a residual constant in Fig.~\ref{fig:screenlaw} is the numerical confirmation that the identification holds.

\section{The oscillator forbidden states and the two-body prescriptions}
\label{app:ho}

The harmonic-oscillator forbidden states used in Sec.~\ref{sec:prescription} are the functions from which the number of forbidden states in the orthogonality-condition model is derived,
\begin{equation}
\mathcal{R}_{nL}(r)=\mathcal{N}_{nL}\,r^{L}L_{n}^{(L+1/2)}(2\nu r^{2})\,e^{-\nu r^{2}},\qquad
\varphi_{nL}(p)=(-1)^{n}\mathcal{N}'_{nL}\,p^{L}L_{n}^{(L+1/2)}\!\Bigl(\frac{p^{2}}{2\nu}\Bigr)e^{-p^{2}/4\nu},
\label{eq:ho}
\end{equation}
with $L_{n}^{(L+1/2)}$ the associated Laguerre polynomials, $n$ the radial quantum number, $\nu=\mu\omega/2\hbar$ the width of the pair, and the momentum-space form following from the self-duality of the oscillator under the Fourier--Bessel transform. Table~\ref{tab:ho} lists the four states used, their widths and their overlaps with the deep forbidden states of the interactions of Sec.~\ref{sec:tables}.

\begin{table}[htbp]
\caption{The oscillator forbidden states of Sec.~\ref{sec:prescription}: pair and wave, radial and momentum-space forms, width, and overlap with the interaction's own forbidden state.}
\label{tab:ho}
\begin{center}
\small
\begin{tabular}{llllcc}
\toprule
pair & state ($n$, $L$) & $\mathcal{R}_{nL}(r)\propto$ & $\varphi_{nL}(p)\propto$ & $\nu$ (fm$^{-2}$) & overlap \\
\midrule
$\alpha N$ ($^{2}S_{1/2}$) & $0s$ (0, 0) & $e^{-\nu r^{2}}$ & $e^{-p^{2}/4\nu}$ & $0.270$\textsuperscript{a} & $0.989$ \\
$\alpha\alpha$ ($L=0$) & $0s$ (0, 0) & $e^{-\nu r^{2}}$ & $e^{-p^{2}/4\nu}$ & $0.5625$ & $0.996$ \\
$\alpha\alpha$ ($L=0$) & $1s$ (1, 0) & $(1-\tfrac{4}{3}\nu r^{2})\,e^{-\nu r^{2}}$ & $(1-p^{2}/3\nu)\,e^{-p^{2}/4\nu}$ & $0.5625$ & $0.983$ \\
$\alpha\alpha$ ($L=2$) & $0d$ (0, 2) & $r^{2}e^{-\nu r^{2}}$ & $p^{2}e^{-p^{2}/4\nu}$ & $0.5625$ & $0.990$ \\
\bottomrule
\end{tabular}
\par\noindent{\footnotesize \textsuperscript{a}~$\nu=1/2b^{2}$ at $b=1.36$~fm.}
\end{center}
\end{table}

\begin{table}[htbp]
\caption{Coefficients $A^{(nL)}_{k}$ of the two-pole form~(\ref{eq:hoff}) of the four oscillator forbidden states of Table~\ref{tab:ho}, in the reduced form $\varphi_{nL}/p^{L}$ of the ancillary tables, with the pole scales in fm$^{-1}$ and the overlap of the form factor with the exact oscillator function. The $\alpha N$ column is that of the $\alpha n$ and the $\alpha p$ channel alike.}
\label{tab:hocoef}
\begin{center}
\scriptsize
\setlength{\tabcolsep}{4pt}
\begin{tabular}{lcccc}
\toprule
 & $\alpha N$ $0s$ & $\alpha\alpha$ $0s$ & $\alpha\alpha$ $1s$ & $\alpha\alpha$ $0d$ \\
\midrule
$\beta$   & $2.0$ & $3.0$ & $3.0$ & $3.0$ \\
$\gamma_{nL}$ & $0.5723023$ & $2.0$ & $3.5$ & $2.5$ \\
$M$       & 1 & 1 & 1 & 2 \\
overlap   & $1.000000$ & $1.000000$ & $1.000000$ & $0.99995$ \\
\midrule
$k$ & \multicolumn{4}{c}{$A^{(nL)}_{k}$} \\
\midrule
0 & $2.5892220456\times10^{0}$ & $1.0977814798\times10^{1}$ & $-1.1213334566\times10^{1}$ & $1.5145269895\times10^{1}$ \\
1 & $3.8476370698\times10^{0}$ & $2.0204997365\times10^{1}$ & $1.5765402196\times10^{1}$ & $2.8895236002\times10^{1}$ \\
2 & $-9.1546768035\times10^{-2}$ & $9.9843918447\times10^{0}$ & $8.3982191739\times10^{1}$ & $1.6335316508\times10^{1}$ \\
3 & $-2.6360284762\times10^{0}$ & $-2.3135593051\times10^{0}$ & $8.2353094861\times10^{1}$ & $-5.4261837690\times10^{-1}$ \\
4 & $-1.2572949281\times10^{0}$ & $-3.5265175933\times10^{0}$ & $1.5656209191\times10^{1}$ & $-4.1289911084\times10^{0}$ \\
5 & $5.1057628746\times10^{-1}$ & $-7.8327259149\times10^{-2}$ & $-2.1305365996\times10^{1}$ & $-4.2926326166\times10^{-1}$ \\
6 & $3.2211166309\times10^{-1}$ & $1.0116932307\times10^{0}$ & $-5.4488496562\times10^{0}$ & $1.0169227385\times10^{0}$ \\
7 & $-1.2963243251\times10^{-1}$ & $-1.5370904275\times10^{-1}$ & $6.1114968077\times10^{0}$ & $-7.6981204501\times10^{-2}$ \\
8 & $-1.0106458814\times10^{-1}$ & $-1.5771331465\times10^{-1}$ & $1.0740437097\times10^{0}$ & $-2.6022992230\times10^{-1}$ \\
9 & $9.1434189329\times10^{-2}$ & $7.4062666806\times10^{-3}$ & $-2.3040079815\times10^{0}$ & $-3.8009698434\times10^{-3}$ \\
10 & $-2.4427847292\times10^{-2}$ & $9.2408168516\times10^{-2}$ & $3.8958014257\times10^{-1}$ & $5.1174671608\times10^{-2}$ \\
11 & -- & $-6.6284415202\times10^{-2}$ & $5.5567944341\times10^{-1}$ & $-7.2083920448\times10^{-2}$ \\
12 & -- & $1.9407441751\times10^{-2}$ & $-3.8957271108\times10^{-1}$ & $-5.8874663216\times10^{-3}$ \\
\bottomrule
\end{tabular}
\end{center}
\end{table}

Each $\varphi_{nL}$ is represented by a form factor of the two-pole form
\begin{equation}
\varphi_{nL}(p)\simeq\frac{p^{L}\sum_{k=0}^{n_{c}}A^{(nL)}_{k}P_{k}(x)}{(\beta^{2}+p^{2})(\gamma_{nL}^{2}+p^{2})},
\label{eq:hoff}
\end{equation}
the form of Eq.~(\ref{eq:fbff}) with $x$ the variable of Eq.~(\ref{eq:estff}) and $\beta$, $\gamma_{nL}$ the two pole scales of the fit, which, unlike the $\gamma_{f}$ of Eq.~(\ref{eq:fbff}), correspond to no bound-state energy: they are chosen for the largest overlap of the fitted form with the oscillator function, the $A^{(nL)}_{k}$, $n_{c}$ being the order of the channel, following by linear least squares. For the $\alpha\alpha$ states the overlap is insensitive to $\gamma_{nL}$ and rounded values are used. The coefficients, pole scales and overlaps are listed in Table~\ref{tab:hocoef}; the overlap of the $0d$ state is limited by the single-pole form of Eq.~(\ref{eq:fbff}) for $M=2$. The form factor replaces the dressed eigenstate in the forbidden block of Eq.~(\ref{eq:basisorder}), with no other change in the calculation.

The $\alpha\alpha$ width is that of the relative motion,  $\nu_{\rm rel}=\mu_{A}\nu_{\alpha}=2\times0.28125$~fm$^{-2}$, the single-$\alpha$ width scaled by the reduced mass number of the relative coordinate; the $\alpha N$ width is $\nu=1/2b^{2}$.
\FloatBarrier
\section{Parameters of the separable potentials}
\label{app:tables}

The separable potentials are constructed by the procedure of Sec.~\ref{sec:tables} from the original interactions of Table~\ref{tab:parents}. The fitted coefficients themselves, the $a^{(L)}_{in}$ of Eq.~(\ref{eq:estff}), the coupling matrices and the $A^{(f)}_{n}$ rows of the forbidden states of Eq.~(\ref{eq:fbff}), are provided in machine-readable form as the ancillary file of Ref.~\cite{NishonovValidation}; the tables below give the quantities derived from them.

\begin{table}[htbp]
\caption{The original interactions represented.}
\label{tab:parents}
\begin{center}
\footnotesize
\resizebox{\textwidth}{!}{%
\begin{tabular}{llll}
\toprule
pair & interaction & form & reduced mass \\
\midrule
$NN$ & CD~Bonn~\cite{Machleidt2001} & published parameters, $nn$, $np$, $pp$, $j\le4$ & $m_{n}/2$, $m_{n}m_{p}/(m_{n}+m_{p})$, $m_{p}/2$ \\
$\alpha N$ & KKNN~\cite{KKNN1979} & published parameters, $^{2}S_{1/2}$, $^{2}P_{3/2}$, $^{2}P_{1/2}$, $^{2}D_{5/2}$, $^{2}D_{3/2}$ & $m_{\alpha}m_{N}/(m_{\alpha}+m_{N})$, $3.8046$~fm$^{-1}$ at the $\alpha p$ mass \\
$\alpha\alpha$ & BFW~\cite{BFW1977} & $-122.6225\,e^{-(0.469\,r)^{2}}$~MeV in every partial wave $+\;4e^{2}\,\mathrm{erf}(0.75\,r)/r$ & $9.518$~fm$^{-1}$ \\
\bottomrule
\end{tabular}}
\end{center}
\end{table}

The quantities new to this paper are those with the Coulomb interaction included, collected in Table~\ref{tab:coulon}: the $pp$ Coulomb-modified threshold parameters, the $\alpha p$ Coulomb-dressed forbidden state and the $^{5}$Li resonance with its Coulomb shift, and the $\alpha\alpha$ Coulomb-dressed forbidden states and the $^{8}$Be resonances.

The resonances are obtained with the screened point Coulomb interaction as in Ref.~\cite{NishonovValidation}: the $^{5}$Li position and width from a Pad\'e approximant of $\det[\bm\Lambda^{-1}-\bm{\mathcal D}(E)]$ in the momentum variable, extrapolated in the screening radius from $50$ to $200$~fm, and the $^{8}$Be states from the $90^{\circ}$ crossings of the screened-Coulomb phase at $80$~fm, the width from the phase slope; the low position of the $^{5}$Li resonance is a property of the KKNN model. The deep $0s$ state of the $L=0$ potential differs by a quarter of an MeV from that of the original potential, a property of the analytic family. The near-threshold behaviour is determined by the support point at the shallow $L=0$ bound state of the original potential without the Coulomb force, whose position the fitted potential reproduces to $34$~keV; with the Coulomb force this state is the $^{8}$Be ground-state resonance of Table~\ref{tab:coulon}. The $^{8}$Be $2^{+}$ and $4^{+}$ entries are the $\delta_{L}=90^{\circ}$ crossings of the phase with the width from its slope; the fitted $4^{+}$ has a numerical uncertainty of $50$~keV.

\begin{table}[htbp]
\caption{Quantities with the point Coulomb interaction included, from the separable potentials (fit) and the original potentials, against experiment (the $pp$ threshold parameters from Ref.~\cite{Bergervoet1988}, the resonances from Refs.~\cite{Tilley2002,Tilley2004}; MeV unless stated; $^{8}$Be energies relative to the $\alpha\alpha$ threshold). For the $\alpha N$ resonances fit and original potential coincide.}
\label{tab:coulon}
\begin{center}
\footnotesize
\begin{tabular}{llccc}
\toprule
pair & quantity & fit & original & experiment \\
\midrule
$pp$ & $a^{C}_{pp}$, $r^{C}_{pp}$ (fm) & $-7.815(2)$, $2.77(7)$ & $-7.8154$, $2.773$ & $-7.8149$, $2.769$ \\
\midrule
$\alpha p$ & forbidden $0s$ & $-10.001$ & $-9.998$ & -- \\
 & $^{5}$Li $3/2^{-}$: $E_{r}$, $\Gamma$ & \multicolumn{2}{c}{$1.55$, $1.12$} & $1.69$, $1.23$ \\
 & $^{5}$He$\to{}^{5}$Li Coulomb shift & \multicolumn{2}{c}{$0.87$} & $0.89$ \\
\midrule
$\alpha\alpha$ & forbidden $0s$, $1s$ & $-72.53$, $-25.87$ & $-72.786$, $-25.879$ & -- \\
 & forbidden $0d$ & $-22.29$ & $-22.290$ & -- \\
 & $^{8}$Be $0^{+}$ & $0.1059$ & $0.0870$ & $0.0918$ \\
 & $^{8}$Be $2^{+}$: $E_{r}$, $\Gamma$ & $3.31$, $2.09$ & $3.32$, $2.11$ & $3.12$, $1.51$ \\
 & $^{8}$Be $4^{+}$: $E_{r}$, $\Gamma$ & $12.45$, $4.12$ & $12.52$, $4.31$ & $11.44$, $3.5$ \\
\bottomrule
\end{tabular}
\end{center}
\end{table}

The construction parameters of every separable potential (rank, support points, $\beta$, $M$, $n_{c}$) and their reproduction of the original interactions without the Coulomb interaction, phase shifts, bound states, threshold parameters and off-shell deviations, are given in Ref.~\cite{NishonovValidation}. The CD~Bonn potential was constructed from its published parameters and reproduces the published phase shifts of every $j\le4$ wave in all three charge states to the rounding of the published tables from $1$ to $300$~MeV. The $\alpha N$ and $\alpha\alpha$ separable potentials reproduce the phase shifts of their original potentials over $1$--$22$~MeV very well~\cite{NishonovValidation}.

\FloatBarrier
\bibliographystyle{unsrt}
{\footnotesize
\bibliography{refs}
}
\end{document}